\documentclass[%
 reprint,
 amsmath,amssymb,
 aps,
]{revtex4-2}

\usepackage{graphicx}% Include figure files
\usepackage{dcolumn}% Align table columns on decimal point
\usepackage{bm}% bold math
\usepackage{url} 
\usepackage{hyperref}

\newcommand{\tot}{{\mathrm{tot}}}
\newcommand{\Btot}{{[B]_\mathrm{tot}}}
\newcommand{\Atot}{{[A]_\mathrm{tot}}}
\newcommand{\Xtot}{{[X]_\mathrm{tot}}}
\newcommand{\Ytot}{{[Y]_\mathrm{tot}}}
\newcommand{\Ctot}{{C_\mathrm{tot}}}
\newcommand{\ATP}{{[\mathrm{ATP}]}}
\newcommand{\GTP}{{[\mathrm{GTP}]}}
\newcommand{\ADP}{{[\mathrm{ADP}]}}
\newcommand{\GDP}{{[\mathrm{GDP}]}}
\newcommand{\kc}{{k_\mathrm{cpl}}}
\newcommand{\cpl}{{\mathrm{cpl}}}
\newcommand{\st}{{\mathrm{st}}}

\newcommand{\Etot}{{E_\mathrm{tot}}}

\newcommand{\D}{{\bf D}}
\newcommand{\R}{{\bf R}}
\newcommand{\rAB}{r}

\usepackage{color}

\begin{document}

\preprint{APS/123-QED}

\title{Thermodynamic cost–controllability tradeoff in metabolic currency coupling}

\author{Jumpei F. Yamagishi}
 \email{jumpei.yamagishi@riken.jp}
 \affiliation{Center for Biosystems Dynamics Research, RIKEN, 6-7-1 Minatojima-minamimachi, Chuo-ku, Kobe 650-0047, Japan}
 \affiliation{Universal Biology Institute, The University of Tokyo, 7-3-1 Hongo, Tokyo 113-0033, Japan}
\author{Tetsuhiro S. Hatakeyama}%
\affiliation{Earth-Life Science Institute, Institute of Future Science, Institute of Science Tokyo, Tokyo 152-8550, Japan}%

\date{\today}% It is always \today, today,

\begin{abstract}
Cellular metabolism is globally regulated by various currency metabolites such as ATP, GTP, and NAD(P)H. These metabolites cycle between charged (high-energy) and uncharged (low-energy) states to mediate energy transfer. While distinct currency metabolites are associated with different metabolic functions, {their charged and uncharged forms are generally interchangeable via biochemical reactions such as ${\rm ATP{\,+\,}GDP{\,\rightleftharpoons\,}ADP{\,+\,}GTP}$ and $\rm NADP^+{\,+\,}NADH{\,\rightleftharpoons\,}NADPH{\,+\,}NAD^+ $.} Thus, their energetic states are generally coupled and influence each other, which would hinder the independent regulation of different currency metabolites. 
    {Despite the extensive knowledge of the molecular biology of individual currency metabolites, it remains poorly understood how the coordination of various coupled currency metabolites shapes metabolic regulation, efficiency, and ultimately the evolution of organisms.} 
    Here, we present a minimal theoretical model of metabolic currency coupling and reveal a fundamental tradeoff relationship between {metabolic controllability and thermodynamic cost: increasing the capacity to independently regulate multiple currency metabolites generally requires comparable abundances of those metabolites, which in turn incurs a higher entropy production rate.} 
    The tradeoff suggests that in complex environments, organisms evolutionarily favor {an equal abundance of currency metabolites} to enhance metabolic controllability at the expense of a higher thermodynamic cost; conversely, in simple environments, organisms evolve to have imbalanced amounts of them to reduce {heat dissipation. These considerations also offer a hypothesis regarding evolutionary trends in nucleotide-pool balance and genomic GC content.}
\end{abstract}

\maketitle

%\tableofcontents

\section{Introduction}

Living organisms must continually consume energy to maintain their order and avoid reaching thermodynamic equilibrium. In the 1940s, Erwin Schr\"odinger famously argued that life persists by avoiding equilibrium through metabolism~\cite{schrodinger1944life,ornes2017nonequilibrium}. Since then, cellular metabolism has remained a central topic in the study of nonequilibrium thermodynamics, attracting the interest of physicists and biologists alike~\cite{von1950theory,von2013biothermodynamics}. 
{Remarkably, recent advancements in stochastic thermodynamics have provided a rigorous nonequilibrium thermodynamic description for open chemical reaction networks, applicable to general metabolic reaction networks~\cite{sekimoto2010stochastic,schmiedl2007stochastic,ge2016mesoscopic,rao2016nonequilibrium,yoshimura2021information,cao2025stochastic}. 
    However, most theoretical efforts to date have taken one of two extremes: they either derive very general thermodynamic constraints applicable to arbitrary chemical reaction networks \cite{schmiedl2007stochastic,rao2016nonequilibrium,yoshimura2021information}, or they analyze specific, simple biochemical processes such as single-enzyme kinetics \cite{seifert2011stochastic,rao2015thermodynamics} and molecular machines \cite{golubeva2012efficiency}.} 
    {A substantial conceptual gap remains between broad theoretical frameworks and detailed, context-specific biochemical studies. To reveal the key biophysical principles of metabolism, we must bridge this gap by developing a coarse-grained theory that preserves the unique characteristics of metabolic systems}. 

{One approach to appropriately coarse-graining the complexity of metabolism is to focus on a special class of compounds that globally mediate energy transfer across many metabolic reactions: \textit{coenzymes}.} 
    A coenzyme is a small, non-protein organic compound that associates loosely with enzymes and participates in enzymatic reactions as a transferable carrier of chemical groups or electrons~\cite{de1997glossary,goldford2022protein}. 
Well-known examples of coenzymes include adenosine triphosphate (ATP), which carries high-energy phosphate groups, and nicotinamide adenine dinucleotide (NADH/NAD$^+$), which carries high-energy electrons. {Coenzymes typically cycle between high-energy (charged) and low-energy (uncharged) forms.} % ~\cite{reich1981energy,hatakeyama2017metabolic,west2023dynamics}. 
ATP, for instance, is converted by hydrolysis into its lower-energy forms ADP or AMP (adenosine diphosphate or monophosphate), releasing energy, whereas ADP can be recharged back into ATP {through processes such as cellular respiration or photosynthesis}.

\begin{figure*}[t]
    \centering \includegraphics[width=0.98\linewidth]{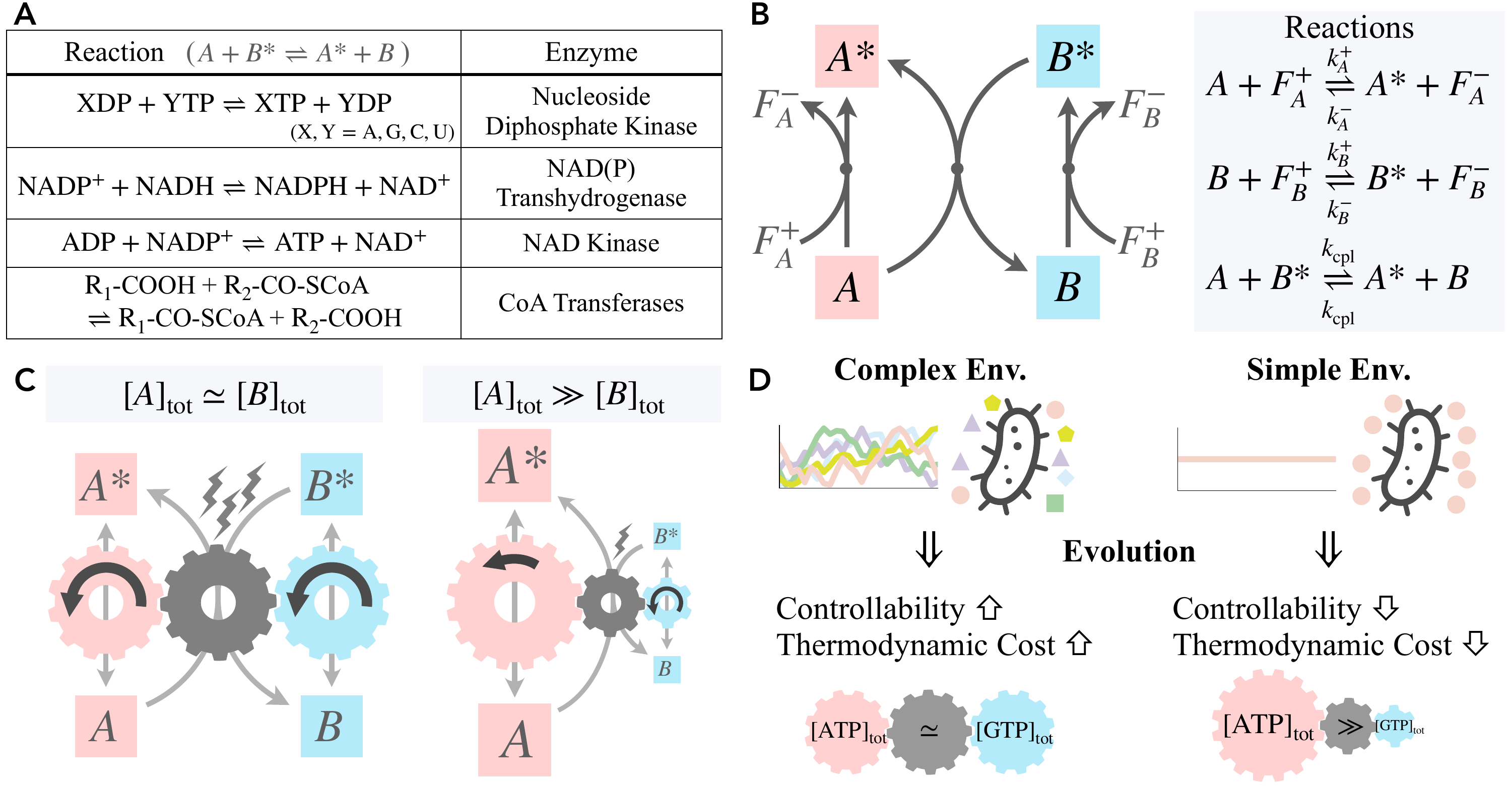}
    \caption{
    Metabolic currency coupling. 
    (A) Examples of transferases and their associated currency coupling reactions. 
    (B) Model of the coupling of currency metabolites. All reactions are reversible. Table~\ref{table:Symbols} collects all the notation in one place. 
    (C) Schematics of how the balance in currency metabolite pools affects metabolic controllability and thermodynamic cost. Coupling reactions cause the ratios $[A^\ast]/[A]$ and $[B^\ast]/[B]$ to shift in the same direction. The former strongly influences the latter when $\Atot\gg\Btot$. Conversely, a large $\Atot/\Btot$ ratio reduces the entropy production rate required to maintain {the metabolic currency coupling, which is analogous to the ``friction'' between different currency metabolites}. 
    (D) A theoretical connection between organismal or habitat complexity and pool sizes of multiple currency metabolites. 
    }
    \label{fig:model}
\end{figure*}

{Because of their central role in energy transfer, coenzymes are often referred to as \textit{currency metabolites} or \textit{metabolic currencies}~\cite{huss2007currency,gerlee2009pathway}. Much like money in an economy, currency metabolites fulfill Jevons’s functions of money~\cite{jevons1875money,mankiwMacroeconomics2013}: they serve as a medium of exchange (transferring energy or chemical groups), a unit of account (providing a characteristic energy metric across reactions), and a store of value (storing energy).} 
The charged/uncharged ratios of currency metabolites, such as the ATP/ADP ratio, % {are considered to} 
regulate metabolic activity globally and {can} serve as macroscopic indicators of the cellular metabolic state~\cite{goldford2022protein,metallo2013understanding,koebmann2002glycolytic,bekiaris2023network}: a higher ratio biases metabolism toward overall ATP-driven reactions over ADP-driven ones, while a lower ratio {favors ADP-driven reactions.} 

Cellular metabolic systems employ multiple different currency metabolites (e.g., ATP, GTP, NAD(P)H, and acetyl-CoA), each generally associated with specific biochemical functions~\cite{bekiaris2023network,west2023dynamics,west2025co}: for example, ATP is known as the primary energy currency for most biochemical reactions, while GTP works for signaling and protein synthesis processes, such as powering G-proteins and the translation of mRNA~\cite{wittinghofer2011structure,mrnjavac2025gtp}; 
    among the electron carriers, NADPH predominantly drives anabolic (biosynthetic) reactions, while NADH feeds into catabolic reactions, such as cellular respiration~\cite{ying2008nad+,canto2015nad+}. 
    {Cells are considered to coordinate various metabolic demands separately by allocating} different currency metabolites to different tasks.

Despite their distinct roles, the energetic states of different currency metabolites are thermodynamically and chemically coupled rather than isolated. 
{In general, the charged and uncharged forms of distinct currency metabolites can be interconverted} through specific enzymatic reactions~\cite{reich1981energy}. 
For example, nucleoside diphosphate kinase catalyzes a reversible phosphate exchange between nucleoside diphosphates and triphosphates (NDPs and NTPs), effectively linking the ATP/ADP and GTP/GDP ratios through the reaction ATP + GDP $\rightleftharpoons$ ADP + GTP \cite{lascu2000catalytic}; other transferases (EC 2) perform analogous exchanges (Fig.~\ref{fig:model}A). Such coupling reactions will hinder the independent regulation of different currency metabolites: if the intracellular ATP/ADP ratio is increased, for instance, coupling reactions will inevitably push the GTP/GDP ratio upward as well. 

While extensive knowledge of the molecular biology of individual currency metabolites has been accumulated, how the coordination of various coupled currency metabolites influences metabolic regulation, efficiency, and ultimately the physiology and evolution of organisms remains poorly understood. 
    To address this gap, we theoretically elucidate inherent, system-wide constraints on how cellular metabolism is regulated via metabolic currency coupling. 

In this study, we develop a minimal theoretical model of metabolic currency coupling. {We thereby derive a fundamental tradeoff between metabolic controllability and thermodynamic cost: balanced currency metabolite pools enhance metabolic control at the expense of increased entropy production rate as a thermodynamic cost (see also Fig.~\ref{fig:model}C)}. 
From a biological standpoint, this tradeoff suggests that {organisms} in complex environments would evolutionarily favor uniformity in currency metabolite pools to increase their metabolic controllability; in contrast, those in simpler environments would evolutionarily favor a skewed distribution of currency pools to reduce the thermodynamic cost (Fig.~\ref{fig:model}D). 
    We then explore biological and evolutionary implications of this tradeoff, including a hypothetical link between ATP–GTP coupling, organismal complexity, and genomic GC content.

\section{Model}\label{sec:model}
As the simplest case, we consider the coupling of two currency metabolites, {$A$ and $B$} (see also Fig.~\ref{fig:model}B and Table~\ref{table:Symbols}). 

Each currency metabolite {is driven} from its uncharged state $X$ into its charged state $X^\ast$ via a single effective reaction {($X{\,=\,}A,B$):
\begin{eqnarray}\label{reac:energy}
    A + F_A^+ 
    \overset{k^+_{A}}{\underset{k^-_{A}}{\rightleftharpoons}} 
    A^\ast + F_A^-
    ,\quad
    B + F_B^+ \overset{k^+_{B}}{\underset{k^-_{B}}{\rightleftharpoons}} B^\ast + F_B^-.
\end{eqnarray}
}Here, $F_X^+$ coarse-grains a pool of energy‐rich {``fuel''} substrates (e.g., nutrients or reducing equivalents), and $F_X^-$ represents a lumped set of lower-energy products or waste. 
{Thus, this reversible reaction represents both catabolic (rightward) and anabolic (leftward) processes.} 

Different currency metabolites interconvert between their charged and uncharged states as 
\begin{eqnarray}\label{reac:exchange}
    A + B^\ast 
    \overset{k^+_\cpl}{\underset{k^-_\cpl}{\rightleftharpoons}} 
    A^\ast + B.
\end{eqnarray}
For the sake of simplicity, it is assumed that the difference in standard chemical potentials between $X^\ast$ and $X$, $\Delta\mu^\circ_X{\,:=\,}\mu_{X^\ast}^\circ - \mu_X^\circ>0$, is identical for both currency metabolites $X=A,B$. From the local detailed balance condition, $k^+_\cpl{\,=\,}k^-_\cpl$ then holds, which will be denoted by $\kc$ hereafter. 

Assuming mass action kinetics, the rate equation for this system is given as 
\begin{eqnarray} \label{eq:model_dynamics}
    \frac{{\rm d}}{{\rm d} t} \begin{pmatrix}
            [A] \\
            [A^\ast] \\
            [B] \\
            [B^\ast]
            \end{pmatrix} &=& \begin{pmatrix}
            -1 & 0 & -1 \\
            1 & 0 & 1 \\
            0 & -1 & 1 \\
            0 & 1 & -1 
            \end{pmatrix}\begin{pmatrix}
            J_A \\
            J_B \\
            J_\cpl 
            \end{pmatrix}, \\
    J_A &=& \kappa^+_A[A] - \kappa^-_A[A^\ast], \nonumber \\ 
    J_B &=& 
    \kappa^+_B[B] -\kappa^-_B[B^\ast], \nonumber \\ 
    J_\cpl &=& 
    \kc[A][B^\ast] - \kc[A^\ast][B]. \nonumber
\end{eqnarray}
{Here,} the effective rate constants {for the driving reactions}, defined as $\kappa_X^\pm{\,:=\,}k_X^\pm [F_X^\pm]$, quantify how strongly currency metabolite $X$ is driven {toward} its charged and uncharged state by catabolic and anabolic processes. {In reality, $\kappa_X^\pm$ can vary over time via transcriptional or translational regulation, while these regulatory processes typically occur on slower timescales than the relaxation of the coupling reactions~\cite{reich1981energy}. In the following analysis, we therefore treat $F_X^+$ as effectively \textit{chemostated} and $\kappa_X^\pm$ as constants.} 

Each reaction conserves {the total pool of each currency metabolite: $\Atot{\,:=\,}[A^\ast]{\,+\,}[A]$ and $\Btot{\,:=\,}[B^\ast]{\,+\,}[B]$} are conserved moieties. 
By exploiting the symmetry between currency metabolites, we order their total pool sizes as $\Atot\,{\geq}\,\Btot$, unless otherwise stated.

\begin{table}[tb]
    \centering \caption{Notations} \label{table:Symbols}
    \begin{tabular}{c|c}
    Symbol & Description  \\ \hline 
    $X,X^\ast$ & Currency metabolite in uncharged/charged state $(X{\,=\,}A{,}B)$ \\
    $[X]_\tot$ & {Pool size} of currency metabolite $X$ \\
    $\kc$ & Rate constant for the coupling reaction (Eq.~\eqref{reac:exchange}) \\
    $k^\pm_X$ & Rate constant for the driving reactions (Eq.~\eqref{reac:energy}) \\
    $F_X^\pm$ & {Fuel ($F_X^+$) and product ($F_X^-$)} for the driving reactions \\
    $\kappa^\pm_X, \kappa_X$ & Effective rate constant for the driving reactions \\
      & ($\kappa_X^\pm{:=}k_X^\pm [F_X^\pm]; \kappa_X{:=}\kappa_X^+{+}\kappa_X^-$) \\
    $\Gamma_X$ & Charged/Uncharged ratio of $X$ ($\Gamma_X{\,:=\,}[X^\ast]/[X]$) \\
    $\Gamma_0$ & $\Gamma_X$ in the strong coupling limit (Eq.~\eqref{eq:gamma_0})\\
    $e^\pm_{XY}$ & Elasticity of $\Gamma_X$ by $\kappa_Y^\pm$ (Eq.~\eqref{eq:elasticity_def}) \\
    $e_X$ & Geometric mean of $e^\pm_{XX}$ (Eq.~\eqref{eq:mean_e}) \\
    $e$ & Mean elasticity (Eq.~\eqref{eq:total_e}) \\
    $\tilde{e}^\pm_{X}$ & $e^\pm_{YX}$ in strong coupling limit \\ 
    $\sigma_\cpl$ & Entropy production rate (EPR) of the coupling reaction  
    \end{tabular}
\end{table}

\section{Results}
{The present study focuses} on the charged/uncharged ratios of currency metabolites $X$, $\Gamma_X{\,:=\,}[X^\ast]/[X]$. These ratios determine the macroscopic state of intracellular metabolism: when $\Gamma_X$ increases (decreases), the net fluxes of $X^\ast$-driven reactions are {preferentially promoted (suppressed)} overall. 

\begin{figure}[tb]
    \centering 
    \includegraphics[width=\linewidth, clip]{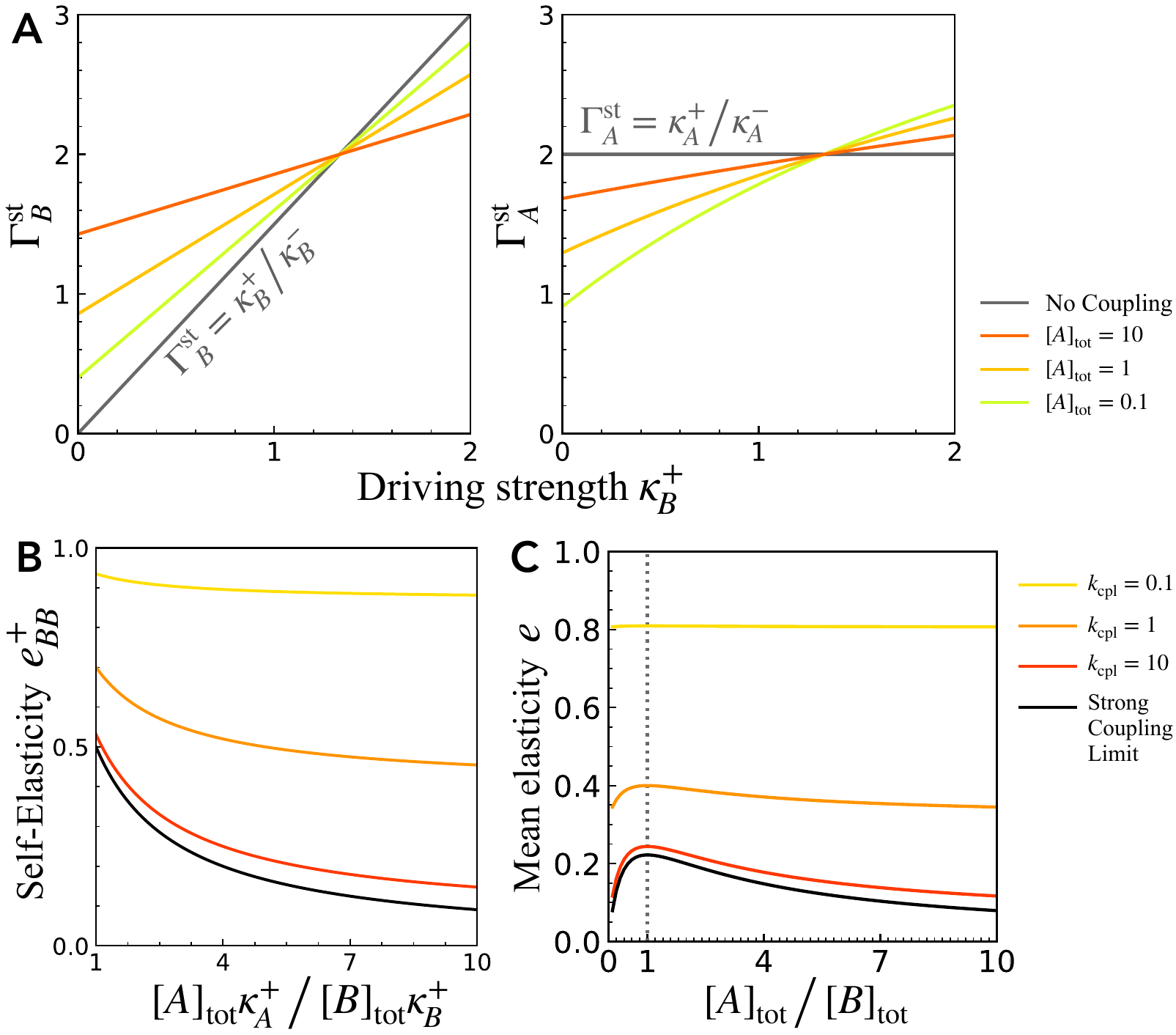} 
    \caption{Elasticity of the charged/uncharged ratios with metabolic currency coupling. 
        (A) Dependence of charged/uncharged ratio $\Gamma_X^\st\,(X{\,=\,}A,B)$ on $\kappa_B^+$. Colored lines show the cases with $\kc{\,=\,}1,\Btot{\,=\,}1$ and different $\Atot$, while gray line shows the no coupling case. 
        (B) Dependence of {self-elasticity} $e^+_{BB}$ on $\Atot\kappa_A^+ / \Btot\kappa_B^+$. 
        Colored lines show different coupling strength $\kc$ {and} black line shows the large coupling limit, $\tilde{e}^+_B$ (Eq.~\eqref{eq:e_XY_k_large}). 
        (C) 
        Relationship between $\Atot/\Btot$ and mean elasticity $e$. The black line shows the strong coupling limit. In panels (B) and (C), $\kappa_B^+{\,=\,}1/3$ and $\Atot{\,+\,}\Btot{\,=\,}2$ are fixed. 
        The other parameters are set as $\kappa_A^+{\,=\,}2/3$, $\kappa_A^-{\,=\,}1/3$, and $\kappa_B^-{\,=\,}2/3$. 
    } \label{fig:controllability}
\end{figure}

First, $\Gamma_X$ at the steady state, denoted by $\Gamma_X^\st$, is calculated as {(see also Appendix~\ref{sec:SM_steady_state})}:
\begin{align}
    & \Gamma_A^\st = \frac{[A^\ast]^\st}{[A]^\st} = 
        \frac{ \kc\Atot\kappa_A^+ + \kc\Btot\kappa_B^+ + \kappa_A^+\kappa_B }
        { \kc\Atot\kappa_A^- + \kc\Btot\kappa_B^- + \kappa_A^-\kappa_B },
    \nonumber \\
    & \Gamma_B^\st = \frac{[B^\ast]^\st}{[B]^\st} = 
    \frac{ \kc\Atot\kappa_A^+ + \kc\Btot\kappa_B^+ + \kappa_A\kappa_B^+ }{ \kc\Atot\kappa_A^- + \kc\Btot\kappa_B^- + \kappa_A\kappa_B^- }, \label{eq:X_ratio} 
\end{align}
where $\kappa_X{\,:=\,}\kappa_X^+{\,+\,}\kappa_X^-$ (see also Fig.~\ref{fig:controllability}A). 

Without currency coupling (i.e., $\kc{\,=\,}0$), {the steady-state ratio satisfies} $\Gamma_X^\st{\,=\,}\kappa_X^+/\kappa_X^-$. {Currency coupling therefore shifts} $\Gamma_X^\st$ from {the ratio of the effective rate constants for the driving reactions,} $\kappa_X^+/\kappa_X^-$, {depending} on the pool size of the other currency metabolite as well as the coupling constant $\kc$. 
In the limit of weak currency coupling, $\kc{\,\ll\,}\kappa_X^\pm/\Xtot$, {we obtain
\begin{align}
    & \Gamma_A^\st = 
    \frac{\kappa_A^+}{\kappa_A^-} + \kc\Btot 
    \frac{ \kappa_B^-}{ \kappa_B\kappa_A^-}
    \left(\frac{\kappa_B^+}{\kappa_B^-}-\frac{\kappa_A^+}{\kappa_A^-}\right) + O(k^2_\cpl), 
    \nonumber \\ 
    & \Gamma_B^\st = 
    \frac{\kappa_B^+}{\kappa_B^-} + \kc\Atot 
    \frac{ \kappa_A^-}{ \kappa_A\kappa_B^-}
    \left(\frac{\kappa_A^+}{\kappa_A^-}-\frac{\kappa_B^+}{\kappa_B^-}\right) + O(k^2_\cpl). 
\end{align}
}$\Gamma_X^\st$ deviates more from ${\kappa_X^+}/{\kappa_X^-}$ toward ${\kappa_Y^+}/{\kappa_Y^-}$ {($Y\,{\neq}\,X$), as $[Y]_\tot$ increases}. 

By contrast, in the strong coupling limit, $\kc\,{\gg}\,\kappa_X^\pm/\Xtot$, both steady‐state ratios {$\Gamma_A^\st$ and $\Gamma_B^\st$} {converge to the identical value:
\begin{eqnarray} 
    \Gamma_0 &:=& \frac{ \Atot\kappa_A^+ + \Btot\kappa_B^+ }{ \Atot\kappa_A^- + \Btot\kappa_B^- } \nonumber \\
    &=& \frac{ (\Atot/\Btot)\kappa_A^+ + \kappa_B^+ }{ (\Atot/\Btot)\kappa_A^- + \kappa_B^- }, \label{eq:gamma_0}
\end{eqnarray}
}which is a weighted average of the individual driving‐force ratios{, $\kappa_A^+/\kappa_A^-$ and $\kappa_B^+/\kappa_B^-$, with weights, $\Atot$ and $\Btot$}. 
{From Eq.~\eqref{eq:gamma_0}, the charged/uncharged ratio with strong currency coupling depends on the currency pool sizes, $\Atot$ and $\Btot$, only through their ratio, $\Atot/\Btot$}.

\subsection{Controllability of charged/uncharged ratios}\label{sec:SS} 
The adaptive regulation of cellular metabolism requires the appropriate modulation of the charged/uncharged ratios $\Gamma_X$ in response to changing environmental conditions such as nutrient availability. 
    To quantify this controllability, we adopt the concept of \textit{elasticity} from economics~\cite{samuelson1947foundations}, measuring the susceptibility of the steady-state charged/uncharged ratios $\Gamma_X^\st$ to variations {in the effective rate constant for a driving reaction, $\kappa_Y^\pm$}: 
\begin{align}
    & e^\pm_{XY} := \pm\frac{\partial \log\Gamma_X^\st}{\partial \log\kappa_Y^\pm}\quad  (X,Y=A,B), 
    \label{eq:elasticity_def}
\end{align}
{where the logarithmic form makes the elasticity dimensionless.} 
As {$\Gamma_X^\st$ ($1/\Gamma_X^\st$)} increases linearly with the driving strength $\kappa_X^+$ ($\kappa_X^-$) as shown in Eq.~\eqref{eq:X_ratio}, each elasticity $e^\pm_{XY}$ takes a {constant, dimensionless} value between $0$ and $1$: {by substituting Eq.~\eqref{eq:X_ratio} into the definition in Eq.~\eqref{eq:elasticity_def} we obtain}
\begin{align}
    e^\pm_{AA} =\;&
    \frac{ \kc\Atot\kappa_A^\pm + \kappa_A^\pm\kappa_B }
    { \kc\Atot\kappa_A^\pm + \kc\Btot\kappa_B^\pm + \kappa_A^\pm\kappa_B }, \nonumber  \\
    e^\pm_{AB} =\;&
    \frac{ \kc\Btot\kappa_B^\pm }
    { \kc\Atot\kappa_A^\pm + \kc\Btot\kappa_B^\pm + \kappa_A^\pm\kappa_B }, 
    \nonumber \\
    e^\pm_{BA} 
    =\;& \frac{ \kc\Atot\kappa_A^\pm }
    { \kc\Atot\kappa_A^\pm + \kc\Btot\kappa_B^\pm + \kappa_A\kappa_B^\pm }, \nonumber \\
    e^\pm_{BB} 
    =\;& \frac{ \kc\Btot\kappa_B^\pm + \kappa_A\kappa_B^\pm }
    { \kc\Atot\kappa_A^\pm + \kc\Btot\kappa_B^\pm + \kappa_A\kappa_B^\pm }. \label{eq:e_BB}
\end{align}
Note that the elasticity of $\Gamma^\st_X$ against different driving reactions satisfies $e^\pm_{XA}{\,+\,}e^\pm_{XB}{\,=\,}1\;(X{\,=\,}A,B)$.

{Equation~\eqref{eq:e_BB} shows} that elasticity is governed not only by {the rate constants for the driving and coupling reactions, $\kappa_X^\pm$ and $\kc$, but also by the currency metabolite pool sizes, $\Xtot$ (Fig.~\ref{fig:controllability}B). In particular, each elasticity {$e^\pm_{YX}$} depends on the relative magnitude of $\Xtot\kappa_X^\pm$.} 
This is particularly evident in the strong coupling limit, $\kc\,{\gg}\,\kappa_X^\pm/\Xtot$: 
\begin{eqnarray}
    e^\pm_{YX} 
    \to \tilde{e}^\pm_X := \frac{ [X]_\tot \kappa_X^\pm }
    { \Atot\kappa_A^\pm + \Btot\kappa_B^\pm}. \label{eq:e_XY_k_large} 
\end{eqnarray}
This $\tilde{e}^\pm_X$ satisfies 
\begin{eqnarray} \label{eq:e_tilde_bounds} 
    e^\pm_{XX} \geq \tilde{e}^\pm_X \geq e^\pm_{YX} \quad (Y\,{\neq}\,X),
\end{eqnarray}
{and it thus provides} the lower (upper) bound of the dimensionless elasticity $e^\pm_{XX}$ ($e^\pm_{YX}$) (see also Fig.~\ref{fig:controllability}B). 

Remarkably, when the currency metabolite pools become imbalanced ($\Atot\,{\gg}\,\Btot$), the less abundant currency $B$ loses susceptibility to its own drives $\kappa_B^\pm$ and becomes predominantly slaved to $\Gamma_A^\st$ and $\kappa_A^\pm$ (Figs.~\ref{fig:controllability}A~and~\ref{fig:controllability}B). In other words, the model effectively behaves as {a single-degree-of-freedom system}. 
Intuitively, this occurs because the more abundant currency metabolite $A$ acts as a {chemical bath} against the less abundant currency metabolite $B$ (Fig.~\ref{fig:model}C).

Taking {the geometric mean of ``self-elasticity'' $e^\pm_{XX}$} yields a single measure of the controllability of each $\Gamma_X^\st$:
\begin{align} \label{eq:mean_e}
    e_X := \sqrt{e^+_{XX} e^-_{XX}} 
    \quad(X=A,B).
\end{align}
Furthermore, the geometric mean of $e_X$ yields a single measure of the {system-wide controllability}:
\begin{align} \label{eq:total_e}
    e := \sqrt{e_A e_B} = \left(e^+_{AA} e^-_{AA} e^+_{BB} e^-_{BB}\right)^{1/4}.
\end{align} 
Consistent with the above discussion, imbalanced currency metabolite pools ($\Atot\,{\gg}\,\Btot$) 
{increase the bias between $e^\pm_{AA}$ and $e^\pm_{BB}$, as well as between $e_A$ and $e_B$,} thereby reducing mean elasticity $e$ (Fig.~\ref{fig:controllability}C). 

{Last but not least, in addition to controllability in terms of elasticity, the achievable ranges of the charged/uncharged ratios $\Gamma_X^\st$ are constrained by the {$\Atot/\Btot$ ratio (Appendix~\ref{sec:bound_modulation_range})}. To achieve the wide range of charged/uncharged ratios observed \textit{in vivo}~\cite{liesa2013mitochondrial}, cells must regulate not only the driving reactions but also the pool sizes of currency metabolites.}

\subsection{Entropy production rate as a thermodynamic cost}\label{sec:transient} % Cellular metabolism requires energy to maintain {its} nonequilibrium state {}. 
{The entropy production rate (EPR) represents a thermodynamic or energetic cost of maintaining nonequilibrium states homeostatically~\cite{himeoka2014entropy,droste2024thermodynamics}, which can be induced by metabolic currency coupling.} 
The EPR associated with coupling reaction, $\sigma_\cpl$, is {defined as}~\cite{yoshimura2021information} 
\begin{align}
    \sigma_\cpl :=\;&\left(\kc[A][B^\ast] - \kc[A^\ast][B]\right) 
        \log\frac{[A][B^\ast]}{[A^\ast][B]} \nonumber \\
    =\;&\kc\Atot\Btot\frac{\Gamma_B - \Gamma_A}{(\Gamma_A+1)(\Gamma_B+1)}\log\frac{\Gamma_B}{\Gamma_A}.
\end{align}
It increases with both the difference between $\Gamma_A$ and $\Gamma_B$ and the product of currency pool sizes, $\kc\Atot\Btot$. Intuitively, the EPR associated with currency coupling can be viewed as the effective ``friction'' between different currency metabolites~{\footnote{{
    In nonequilibrium thermodynamics, friction is a typical irreversible process that produces entropy through heat dissipation. The amount of heat dissipated by mechanical friction increases with distance traveled and the object's weight (more precisely, normal force). 
    In our model, the entropy production rate associated with metabolic currency coupling increases when the charged/uncharged ratios of currency metabolites differ more significantly and when their pool sizes are larger: the former and latter are analogous to the distance traveled and the object's weight or size in mechanical friction, respectively.}}} (Fig.~\ref{fig:model}C). 

\begin{figure}[bt]
    \centering 
    \includegraphics[width=0.99\linewidth, clip]{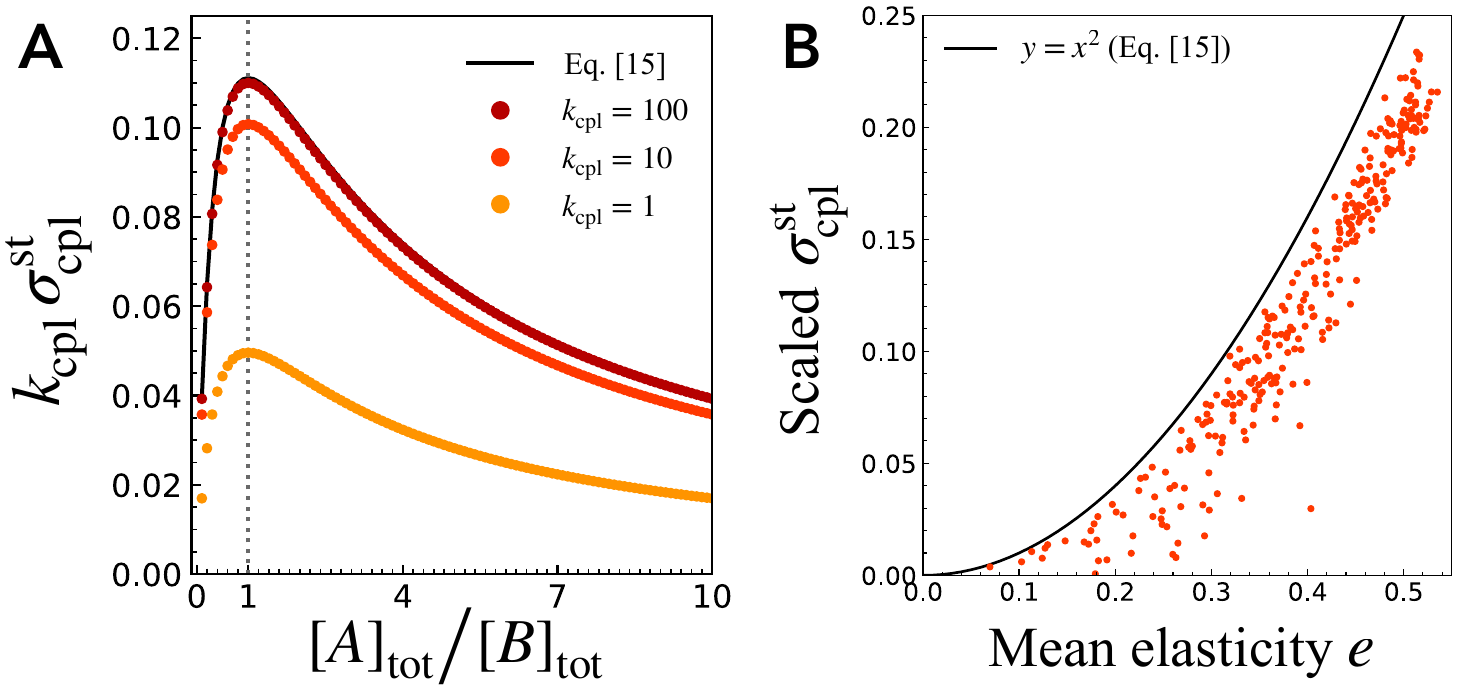}
    \caption{Thermodynamic cost of currency coupling. 
        (A) Dependence of $\kc\sigma_\cpl^\st$ on $\Atot/\Btot$ with relatively strong currency coupling. 
        The black curve represents the strong coupling limit, Eq.~\eqref{eq:EPR_approx_largek}.
        For a consistent comparison, the system size is kept constant by setting $\Atot{\,+\,}\Btot{\,=\,}2$ while $\Atot/\Btot$ is varied; parameters are set as $\kappa_A^+{\,=\,}2/3,\kappa_A^-{\,=\,}1/3$, $\kappa_B^+{\,=\,}1/3,\kappa_B^-{\,=\,}2/3$. 
        (B) Relationship between the scaled EPR of the coupling reaction, $\frac{\kc \sqrt{\kappa_A^+\kappa_A^- \kappa_B^+\kappa_B^-} }{(\kappa_A^-\kappa_B^+ - \kappa_A^+\kappa_B^-)^2} \sigma^\st_\cpl$, and mean elasticity $e$. $\kc{\,=\,}10$. {The parameters $\kappa_X^\pm$ and $\Xtot$ are randomly sampled from a uniform distribution within the range $[0,1]$. The black curve represents the strong coupling limit, Eq.~\eqref{eq:tradeoff_k_large2}.} 
    } \label{fig:EPR}
\end{figure}

Examining the steady-state (housekeeping) EPR $\sigma_{\cpl}^\st$ {in the strong coupling limit ($\kc\,{\gg}\,\kappa_X^\pm/\Xtot$)} provides further {insight} into the thermodynamic cost of currency coupling. 
{In this limit,} the housekeeping EPR of currency coupling is given by (see Appendix~\ref{sec:SM_steady_state} for derivation): 
\begin{align} \label{eq:EPR_approx_largek}
    \sigma^\st_\cpl \simeq\;& \frac{1}{\kc}
    \frac{ (\kappa_A^-\kappa_B^+ - \kappa_A^+\kappa_B^-)^2\Atot \Btot }{ (\Atot\kappa_A^+ + \Btot\kappa_B^+)(\Atot\kappa_A^- + \Btot\kappa_B^-) } 
    \nonumber\\
    \propto\;& \frac{1}{\kc}\frac{\frac{\Atot}{\Btot}}{ \left(\frac{\Atot}{\Btot}\kappa_A^+ + \kappa_B^+\right)\left(\frac{\Atot}{\Btot}\kappa_A^- + \kappa_B^-\right)},
\end{align}
which depends on the $\Atot/\Btot$ ratio, but is independent of {the total abundance of currency metabolites, $\Atot{\,+\,}\Btot$.} 
It is maximized at {$\Atot/\Btot{\,=\,}\left.\sqrt{\kappa_B^+\kappa_B^-} \middle/ \sqrt{\kappa_A^+ \kappa_A^-}\right.$ and decreases as the system deviates from this balanced $\Atot/\Btot$ ratio (Fig.~\ref{fig:EPR}A).} 

A similar approximation-based analysis also applies to the weak coupling limit {($\kc{\,\ll\,}\kappa_X^\pm/\Xtot$)}, yielding a qualitatively consistent result: EPR $\sigma_{\cpl}^\st$ increases as the $\Atot/\Btot$ ratio approaches $1$ (see Appendix~\ref{sec:SM_steady_state} for details).

\subsection{Tradeoff between elasticity and thermodynamic cost}\label{sec:tradeoff}
With strong currency coupling, from Eqs.~\eqref{eq:e_XY_k_large}~and~\eqref{eq:EPR_approx_largek}, the housekeeping EPR $\sigma^\st_\cpl$ and elasticity are related as: 
\begin{align} \label{eq:tradeoff_k_large2}
    & \sigma^\st_\cpl \simeq \frac{1}{\kc}
    \frac{(\kappa_A^-\kappa_B^+ - \kappa_A^+\kappa_B^-)^2}{\sqrt{\kappa_A^+\kappa_A^- \kappa_B^+\kappa_B^- }} 
    e^2.
\end{align}
{This relation} shows a tradeoff {in which} a higher (lower) mean elasticity is associated with a larger (smaller) EPR. 
We also numerically confirmed a positive correlation between $e$ and $\sigma^\st_\cpl$ even outside the strong coupling limit (Fig.~\ref{fig:EPR}B); thus, this tradeoff could generally appear in systems with a coupling of different currency metabolites. 
In addition, the following tradeoff inequality holds for arbitrary $\kc$ (see Appendix~\ref{sec:SM_steady_state} for the derivation):
\begin{align} \label{eq:tradeoff_inequality}
    \frac{(\kappa_A^-\kappa_B^+ - \kappa_A^+\kappa_B^-)^2}{\kc\sqrt{\kappa_A^+\kappa_A^- \kappa_B^+\kappa_B^- }} e^2 \geq \sigma^\st_\cpl .
\end{align}

The tradeoff suggests that greater uniformity in the currency metabolite pools could enhance the controllability of cellular metabolism {and thus increase the} effective degrees of freedom, at the expense of a higher thermodynamic cost (Fig.~\ref{fig:model}C). 
Therefore, organisms inhabiting more complex environments, {which require adaptive regulation of metabolic functions in response to higher-dimensional environmental cues, e.g., independent regulation of glycolysis via ATP, citric acid cycle via GTP, and pentose phosphate pathway via NADPH~\cite{west2025co}}, are expected to favor more balanced currency pools during evolution and adaptation (Fig.~\ref{fig:model}D). 

\subsection{Influence of currency coupling on evolution of GC content}\label{sec:GC}
{The above tradeoff has various biological implications}, including a possible link between ATP–GTP coupling and an evolutionary trend in genomic GC content {(see also Fig.~\ref{fig:model}D)}. 

Genomic GC content refers to the percentage of guanine (G) and cytosine (C) in a genome. Interestingly, the GC content of bacterial genomes is highly diverse, ranging from $15\%$ to $75\%$ across species~\cite{rocha2002base,madin2020synthesis}. 
{Although} several factors have been proposed as contributing to this diversity, such as the differential stability due to the number of hydrogen bonds and sugar-acid preference arising from codon usage bias~\cite{sueoka1961correlation,gralka2023genome}, the origin of this diversity remains unknown.

\begin{figure}[tb]
    % \centering \includegraphics[width=0.875\linewidth, clip]{Fig4.pdf} 
    \centering \includegraphics[width=0.95\linewidth, clip]{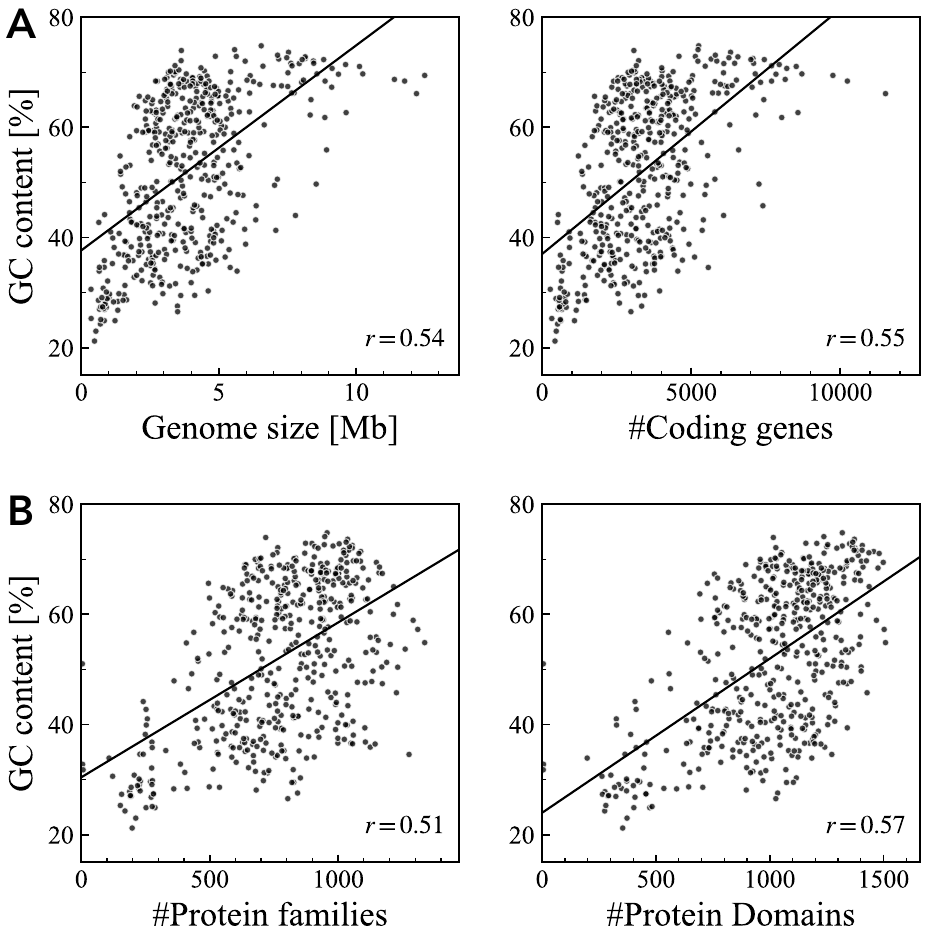} 
    \caption{
        {Statistical} relationship between organismal complexity and GC content. 
        {(A) Correlation between genomic properties (genome size or the number of coding genes) and GC content. 
        (B) Correlation between proteomic properties (the number of protein families or domains) and GC content.} 
        {In (A) and (B), the} data are obtained from dataset~\cite{madin2020synthesis} and Pfam~\cite{mistry2021pfam} for $496$ bacteria and archaea. 
    } \label{fig:bio_data}
\end{figure}

\begin{figure}[tb]
    \centering \includegraphics[width=0.85\linewidth, clip]{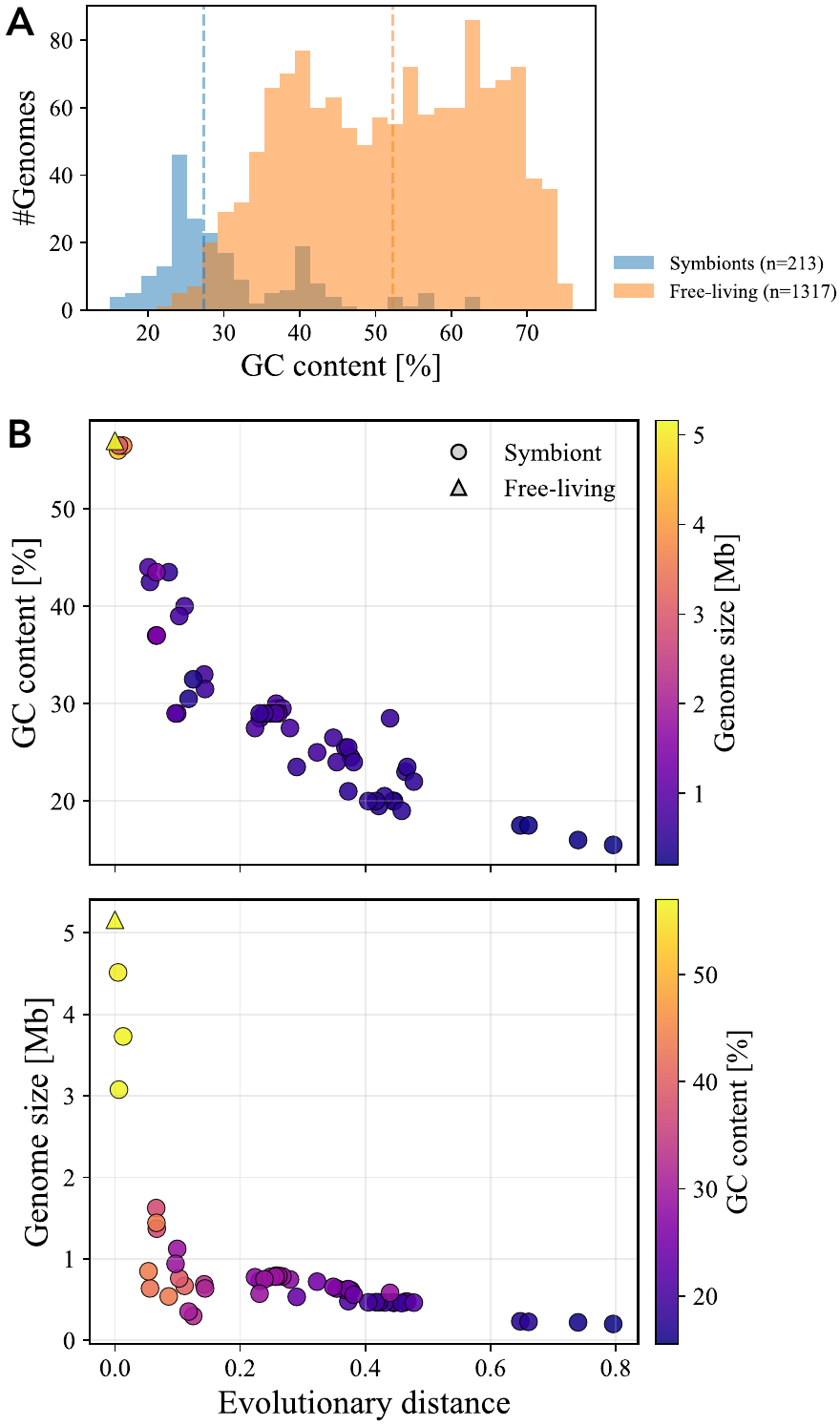} 
    \caption{
        GC-content decline and genome reduction in the evolution of insect endosymbionts. 
        (A) Distribution of genomic GC content in insect endosymbionts (blue) and their free-living relatives (orange). The vertical dotted lines indicate the corresponding medians: $27.4\%$ for symbionts and $52.3\%$ for free-living relatives. Genomes with available FASTA files from the NCBI database are analyzed from the $267$ symbionts and $1317$ free-living relatives compiled in Refs.~\cite{kaltenpoth2025origin,mccutcheon2019life}. 
        (B) Evolutionary distances of \textit{Sodalis}-related insect endosymbionts, defined as tree branch lengths between symbiont lineages and their nearest free-living relative (\textit{Sodalis praecaptivus}), are plotted against genome size (top) and GC content (bottom). Among the symbionts for (A), we extracted $55$ symbiont genomes belonging to the same clade as the genera \textit{Sodalis} and \textit{Buchnera} in the GTDB bacterial reference tree~\cite{Parks2022GTDB,Parks2018GTDBBacteria} (downloaded from \url{https://data.gtdb.ecogenomic.org/releases/release232/232.0/bac120_r232.tree}).
    }\label{fig:evo_dist}
\end{figure}

{Here, we focus on the statistical relationship between} genomic GC content and environmental complexity: 
GC content generally correlates with various indicators of organismal complexity (Fig.~\ref{fig:bio_data}). Fig.~\ref{fig:bio_data}A shows a positive correlation between genomic GC content and the genome size or the number of coding genes---{widely-accepted indicators of organismal complexity, or the required complexity of physiological regulation}~\cite{nelson2004regulatory,sharov2006genome}---across diverse microorganisms{, as previously reported~\cite{almpanis2018correlation}}. GC content also positively correlates with the number of protein families or domains, a recently proposed quantitative indicator of organismal complexity~\cite{alvarez2025organismal} (Fig.~\ref{fig:bio_data}B). 
{More directly, insect endosymbionts, which inhabit relatively simple and stable host-associated environments, exhibit significantly lower GC content than their free-living relatives (Fig.~\ref{fig:evo_dist}A).} 

{From an evolutionary perspective, it has been reported} that organisms evolve to become GC-poor and AT-rich under simple environmental conditions~\cite{foerstner2005environments,almpanis2018correlation}. Endosymbiotic organisms, which inhabit more stable and simplified environments compared to free-living species, evolve to become GC-poor~{\cite{rocha2002base,mccutcheon2019life,mccutcheon2012extreme}}; similarly, mitochondria and chloroplasts typically have GC-poor genomes~\cite{zhang2012complete}. 
Notably, {a phylogenetic analysis of insect endosymbionts in the genus \textit{Sodalis} reported that an initial sharp reduction in genome size due to gene loss is followed by a progressive decrease in GC content~\cite{mccutcheon2019life}; in this regime, genome size and GC content are no longer correlated, but GC content continues to decrease as adaptation to relatively simple host-associated habitats proceeds. 
We found that a similar trend persists even when using updated data covering a broader range of insect endosymbionts (Fig.~\ref{fig:evo_dist}B). % This trend is also observed in  different phylogenetic lineages (Fig.~\ref{fig:evo_dist_SI}). 
These observations are consistent with the possibility that simple host-associated environments evolutionarily favor lower GC content through mechanisms that are independent of genome-size reduction.} 

Our theory on metabolic currency coupling provides a {possible explanation for} the evolutionary tendency of {microbes in simple environments to become GC-poor}. 
{NTPs are the building blocks of RNA, which is the second most abundant substance in cells after proteins~\cite{milo2015cell}. Therefore, the intracellular ATP/GTP ratio may be positively correlated} with genomic GC content~\cite{dietel2019selective} (see also Appendix~\ref{sec:GTP-GC}). 
From the perspective of the above controllability–cost tradeoff due to currency coupling, GC-rich microbes should tend to have higher metabolic controllability but higher thermodynamic cost, and \textit{vice versa}. Although GC content may be affected by many other factors, this suggests that microbes in complex habitats should evolve to increase their GC content to enhance metabolic controllability despite the added energetic burden, while microbes in simple habitats should reduce their GC content to lower thermodynamic cost at the expense of controllability (see also Fig.~\ref{fig:model}D).

\section{Discussion}\label{sec:Discussion}
This work moves beyond the na\"ive view that distinct currency metabolites simply regulate distinct functions, instead providing a theoretical basis that quantitatively links the abundances and distributions of multiple currency metabolites to their physiological roles in managing controllability and energetic cost. 
By developing a minimal model of metabolic currency coupling that does not rely on specific molecules or reactions, we uncover a fundamental tradeoff between the metabolic controllability and the thermodynamic cost: greater controllability of cellular energetic states demands a higher thermodynamic cost, and \textit{vice versa}. 
 
This theory further applies to diverse contexts in {both evolution and adaptation. As an evolutionary example}, we have argued how metabolic currency coupling could influence the evolutionary trend toward more GC-rich microbes in more complex environments {(Fig.~\ref{fig:bio_data}). Comparisons between insect endosymbionts and their free-living relatives are consistent with this hypothesis: endosymbionts, which inhabit relatively stable host-associated environments, tend to evolve toward lower genomic GC content (Fig.~\ref{fig:evo_dist}A). 
In particular, the phylogenetic analysis of insect endosymbionts suggests that GC-content reduction can continue even after substantial genome-size reduction has occurred, implying that the reduction in GC content is not merely a by-product of genome reduction (Fig.~\ref{fig:evo_dist}B). This trend was first reported in a previous study (see Fig.~3 of Ref.~\cite{mccutcheon2019life}), while we show that the same trend is recapitulated using an expanded set of insect endosymbionts and a phylogenetic tree inferred by a different procedure~\footnote{
    {The GTDB reference tree used here (\url{bac120_r232.tree}) is a broad bacterial reference tree inferred with FastTree~\cite{price2010fasttree} from $120$ ubiquitous bacterial marker proteins and one representative genome per GTDB species cluster. In contrast, the phylogeny in Ref.~\cite{mccutcheon2019life} was reconstructed specifically for \textit{Enterobacteriaceae} using RAxML~\cite{stamatakis2014raxml} from $130$ concatenated single-copy orthologs across $93$ selected taxa. Thus, the two analyses differ in taxon sampling, marker gene set, and tree inference method.}}, suggesting the robustness of this evolutionary trend.} 
% This pattern is not restricted to a single clade: similar shifts persist across a broader set of symbiotic microbes and their free-living relatives, although systematic one-to-one comparisons can be confounded by phylogenetic and ecological heterogeneity (Fig.~\ref{fig:bio_data_Appendix}).} 

At the level of non-genetic adaptation, our theoretical results could {also} be experimentally tested by manipulating the pool size of a currency metabolite: e.g., the addition of guanosine, which is converted to GTP via the salvage pathway, can increase GTP levels~\cite{kriel2012direct}. 
Note here that, although our minimal model assumes a constant coupling strength $\kc$ for simplicity, parameter values, including $\kc$, in reality vary dynamically, depending on enzyme and metabolite abundances or growth conditions. An order-of-magnitude estimation of the parameters in our model suggests that $\kc$ and $\kappa^+_{\rm ATP}/\ATP_\tot$ are typically of comparable magnitude, both on the order of $10^3\text{--}10^4\;{\rm [M^{-1}s^{-1}]}$ under physiological conditions (see Appendix~\ref{sec:order} for details); this, in turn, implies that both strong and weak currency-coupling regimes can arise in natural cells as well as in experimentally manipulated systems. 

{For analytical tractability, our model assumes local detailed balance and simple mass-action kinetics for the coupling reaction (Eq.~\eqref{eq:model_dynamics}), as well as symmetry between currency metabolites $\Delta\mu^\circ_A=\Delta\mu^\circ_B$. 
However, our conclusions, including the cost–controllability tradeoff, persist even when we replace mass action by more general reversible enzyme-mediated kinetics (Appendix~\ref{sec:GeneralizedCoupling}.1) or allow asymmetric coupling energetics or small dissipation in currency coupling (Appendix~\ref{sec:GeneralizedCoupling}.2).} 

Our minimal model {also coarse-grains} substrates and anabolic products into a single {quantity, $[F_X^+]$,} to capture the essence of metabolic currency coupling. This leads to the unrealistic feature that the total EPR {vanishes} in the case without currency coupling ($\kc{\,=\,}0$). {In addition, we treat $\kappa_X^\pm$ as constant parameters, reasonably assuming a separation of timescales in which coupling reactions relax sufficiently fast compared to other metabolic processes, particularly the rate-limiting steps in catabolism and anabolism.} 
Such limitations could be resolved by introducing a more detailed and realistic model which incorporates the dynamic change in coupling strength $\kc$ and distinguishes between catabolism and anabolism, without altering our core findings. {For example,} the present model can be easily extended to systems with the coupling of $n{\,\geq\,}3$ currency metabolites, where a similar tradeoff between elasticity and EPR {emerges} (Appendix~\ref{sec:n=3}). 

{The present study introduces the elasticity of the charged/uncharged ratios $e_{XY}^\pm$ (Eq.~\eqref{eq:elasticity_def}) as a natural measure of metabolic controllability, based on economic theory~\cite{samuelson1947foundations}. This quantity is also closely related to concepts in metabolic control analysis (MCA)~\cite{Fell2021MCA,Heinrich1996MetabolicControlAnalysis,fell1985metabolic} (Appendix~\ref{sec:MCA}). 
An interesting future direction is to establish further connections between classical MCA theorems and the concepts in our theory and nonequilibrium thermodynamics, particularly system-wide controllability ($e_X$ and $e$), EPR, and the cost–controllability tradeoff.}

The present study theoretically examines the monetary economics of energy metabolism, thereby revealing a {key tradeoff between metabolic controllability and thermodynamic cost.} 
{As viewing biology through the lens of economics has already yielded valuable insight~\cite{EPCP2025,yamagishi2021microeconomics,yamagishi2023linear,yamagishi2025global}, we anticipate further intersections between cellular metabolism and monetary economics. For example, the economic theory and observation that a new denomination of currency emerges to reduce temporal and spatial {``friction''} in the economy~\cite{kuroda2020global} seems consistent with the recent theory that the redundancy of currency metabolites (coenzymes) facilitates a parsimonious usage of the proteome~\cite{goldford2022protein}. 
We anticipate that further integrating nonequilibrium chemical thermodynamics, systems biology of metabolism, and mathematical economics will uncover universal constraints and design principles governing the economy of cellular metabolism.}

\begin{acknowledgments}
    We would like to thank Ryuna Nagayama for his careful reading of the manuscript. 
    We would also like to acknowledge Kohei Yoshimura, Yuichi Wakamoto, Ken-ichiro F. Kamei, Naoki Konno, {Sosuke Ito, Masaki Fujiyoshi,} and Takuma \={O}nishi for helpful discussions and useful comments. 
    J.~F.~Y. is supported by the RIKEN Research Fund for Special Postdoctoral Researcher (Project Code: 202501094096) and JST ACT-X Grant Number JPMJAX25LG. T.~S.~H. is supported by JSPS KAKENHI Grant Number {JP21K15048. J.~F.~Y. and T.~S.~H. are also supported by JSPS KAKENHI Grant Number JP26K00061.}
\end{acknowledgments}

\appendix
% \renewcommand{\thesubsection}{\Alph{section}.\arabic{subsection}}
% \setcounter{figure}{0}
% \renewcommand{\thefigure}{A\arabic{figure}}

% \begin{figure}[tb]
%     \centering \includegraphics[width=0.8\linewidth, clip]{Fig5_SI_Bacteroidota.pdf} 
%     \caption{
%         {GC-content decline and genome reduction in the evolution of \textit{Chryseobacterium}-related insect endosymbionts. 
%         Evolutionary distances of insect endosymbionts, defined as tree branch lengths between symbiont lineages and their nearest free-living relative (\textit{Chryseobacterium nepalense}), are plotted against genome size (top) and GC content (bottom). Among the insect symbionts listed in Refs.~\cite{kaltenpoth2025origin,mccutcheon2019life}, we extracted $21$ symbiont genomes belonging to the same clade as the order \textit{Flavobacteriales} in the GTDB bacterial reference tree~\cite{Parks2022GTDB,Parks2018GTDBBacteria}. See also Fig.~\ref{fig:evo_dist}. 
%     }} \label{fig:evo_dist_SI}
% \end{figure}

\section{{Derivation of steady-state properties}} \label{sec:SM_steady_state}
The steady state of the present model is calculated as
\begin{align}
    [A^\ast]^\st 
        =\;& \frac{ \kc\Atot\kappa^+_A + \kc\Btot\kappa^+_B + \kappa^+_A\kappa_B }
        { \kc\Atot\kappa_A + \kc\Btot\kappa_B + \kappa_A\kappa_B } \Atot, \nonumber \\ 
    [A]^\st 
        =\;& \frac{ \kc\Atot\kappa^-_A + \kc\Btot\kappa^-_B + \kappa^-_A\kappa_B }
        { \kc\Atot\kappa_A + \kc\Btot\kappa_B + \kappa_A\kappa_B } \Atot, \nonumber \\
    [B^\ast]^\st 
        =\;& \frac{ \kc\Atot\kappa^+_A + \kc\Btot\kappa^+_B + \kappa_A\kappa^+_B }{ \kc\Atot\kappa_A + \kc\Btot\kappa_B + \kappa_A\kappa_B } \Btot, \nonumber \\
    [B]^\st 
        =\;& \frac{ \kc\Atot\kappa^-_A + \kc\Btot\kappa^-_B + \kappa_A\kappa^-_B }{ \kc\Atot\kappa_A + \kc\Btot\kappa_B + \kappa_A\kappa_B } \Btot. 
        \label{eq:SI_SS}
\end{align}
{From Eq.~\eqref{eq:SI_SS}, we can calculate the steady-state charged/uncharged ratios of currency metabolites, $\Gamma_A^\st$ and $\Gamma_B^\st$ (Eq.~\eqref{eq:X_ratio} in the main text). Eq.~\eqref{eq:X_ratio} shows that} $\Gamma_X^\st{\,:=\,}[X^\ast]^\st/[X]^\st$ {and $1/\Gamma_X^\st$} increase linearly with $\kappa_X^+$ {and $\kappa_X^-$, respectively}. 

% In addition, as $\Gamma_X^\st := [X^\ast]^\st / [X]^\st $ is defined, 
% \begin{eqnarray*}
%     [X^\ast]^\st = \frac{ \Gamma_X^\st }{\Gamma_X^\st + 1}\Xtot,\quad [X]^\st = \frac{ 1 }{\Gamma_X^\st + 1}\Xtot.
% \end{eqnarray*}

The entropy production rate (EPR) of the driving reaction for currency metabolite $X{\,=\,}A,B$ is given as follows~\cite{yoshimura2021information}: 
\begin{align*}
    \sigma_X :=\;& \left(\kappa^-_X{[X^\ast]} - \kappa^+_X{[X]} \right)\log\frac{\kappa^-_X{[X^\ast]}}{\kappa^+_X{[X]}} 
        \\ =\;& 
        \frac{\Xtot\kappa^-_X}{\Gamma_X+1}
        \left(\Gamma_X - \Gamma^\kappa_X\right)
        \log\frac{\Gamma_X}{\Gamma^\kappa_X},
\end{align*}
where $ \Gamma^\kappa_X{\,:=\,}\kappa_X^+/\kappa_X^-$. 
That is, the entropy production due to the driving reaction for $X$ arises from the difference between the charged/uncharged ratios $\Gamma_X$ of currency metabolite $X$ and the bias $\Gamma^\kappa_X{\,=\,}\kappa_X^+/\kappa_X^-$ in the effective rate constants of forward and reverse driving reactions.

\noindent\textbf{Weak currency coupling limit.} 
In the limit of weak currency coupling ($\kc{\,\ll\,}\kappa_X^\pm/\Xtot$), the present model is reduced to a linear dynamical system. The steady state, {Eq.~}\eqref{eq:SI_SS}, is then approximated as:
\begin{align*}
    [A^\ast]^\st \simeq\;& 
    \Atot\left(\frac{\kappa_A^+}{\kappa_A} + \kc\frac{\Btot}{\kappa_A}\left( \frac{\kappa_B^+}{\kappa_B} - \frac{\kappa_A^+}{\kappa_A}\right)\right), \\
    [A]^\st \simeq\;&
    \Atot\left(\frac{\kappa_A^-}{\kappa_A} + \kc\frac{\Btot}{\kappa_A}\left( \frac{\kappa_B^-}{\kappa_B} - \frac{\kappa_A^-}{\kappa_A}\right)\right), \\ 
    [B^\ast]^\st \simeq\;& 
    \Btot\left(\frac{\kappa_B^+}{\kappa_B} + \kc\frac{\Atot}{\kappa_B}\left( \frac{\kappa_A^+}{\kappa_A} - \frac{\kappa_B^+}{\kappa_B}\right)\right), \\
    [B]^\st \simeq\;& 
    \Btot\left(\frac{\kappa_B^-}{\kappa_B} + \kc\frac{\Atot}{\kappa_B}\left( \frac{\kappa_A^-}{\kappa_A} - \frac{\kappa_B^-}{\kappa_B}\right)\right).
\end{align*}
Notably, the leading-order behavior of $[X^\ast]^\st/\Xtot$ and $[X]^\st/\Xtot$ exhibits no explicit dependence on $\Xtot$. These are primarily governed by the pool size of the other currency metabolite, $\Ytot\,(Y{\,\neq\,}X)$, with the effect of $\Xtot$ appearing only at second order. 

{The same holds for the charged/uncharged ratio and self-elasticity: e.g., in the case of currency metabolite $B$, 
\begin{align*}
    \Gamma_B^\st \simeq\;& 
    \Gamma^\kappa_B + \kc\Atot \frac{\Gamma^\kappa_B+1}{\Gamma^\kappa_A+1}(\Gamma^\kappa_A-\Gamma^\kappa_B), \\
    e^\pm_{BB} 
    =\;& 1 - \kc\frac{ \Atot\kappa_A^\pm }{\kappa_A\kappa_B^\pm } +O(k_\cpl^2). 
\end{align*}
This indicates that $\Gamma_B^\st$ is determined by a balance between the effective rate constants, $\kappa_B^\pm$, of the driving reaction for currency metabolite $B$ and the driving strength, $\kc\Atot$, of the coupling reaction by the other currency metabolite $A$.}

As a result, $\sigma^\st_i$ ($i=A,B,\cpl$) are approximated as:
\begin{align}
    & \sigma^\st_A \simeq 
    k_\cpl^2\Atot\frac{\kappa_A^+\kappa_A^-}{\kappa_A} \left(\frac{\Btot}{\kappa_B}\left(\frac{ \kappa_B^+ }{\kappa_A^+ } - \frac{ \kappa_B^- }{\kappa_A^- } \right)\right)^2,  \nonumber \\
    & \sigma^\st_B \simeq 
    k_\cpl^2\Btot\frac{\kappa_B^+\kappa_B^-}{\kappa_B} \left(\frac{\Atot}{\kappa_A}\left(\frac{ \kappa_A^+ }{\kappa_B^+ } - \frac{ \kappa_A^- }{\kappa_B^- } \right)\right)^2 ,  \nonumber \\
    & \sigma^\st_\cpl 
    \simeq 
    \kc\Atot\Btot\frac{\kappa^-_A \kappa^+_B - \kappa_A^+ \kappa_B^-}{\kappa_A \kappa_B} \log\frac{\kappa^-_A \kappa^+_B}{\kappa^+_A \kappa^-_B}.  \label{eq:EPR_approx1}
\end{align}
The term in  {Eq.~}\eqref{eq:EPR_approx1}, $\frac{\kappa_A^-\kappa_B^+ - \kappa_A^+ \kappa_B^-}{\kappa_A \kappa_B}\log\frac{\kappa_A^- \kappa_B^+}{\kappa_A^+ \kappa_B^-}$, {is non-negative and} quantifies the extent of asymmetry between the effective rate constants of driving reactions, $\kappa_A^\pm$ and $\kappa_B^\pm$. 
When normalized by the square of the total currency pool $\Ctot{\,:=\,}\Atot{\,+\,}\Btot$, the housekeeping EPR of currency coupling, $\sigma_{\cpl}^\st / \Ctot^2 \propto \kc\rAB/(1+\rAB)^2$ with $\rAB{\,:=\,}\Atot/\Btot$, is maximized at $\rAB=1$. 
Note here that the total EPR, defined as $\sigma_\tot{\,:=\,}\sum_{i=A,B,\cpl}\sigma_i$, satisfies $\sigma^\st_\tot=\sigma^\st_\cpl+O(k_\cpl^2)$ and is primarily governed by the coupling reaction. 

\begin{figure}[tb]
    \centering 
    \includegraphics[width=0.65\linewidth, clip]{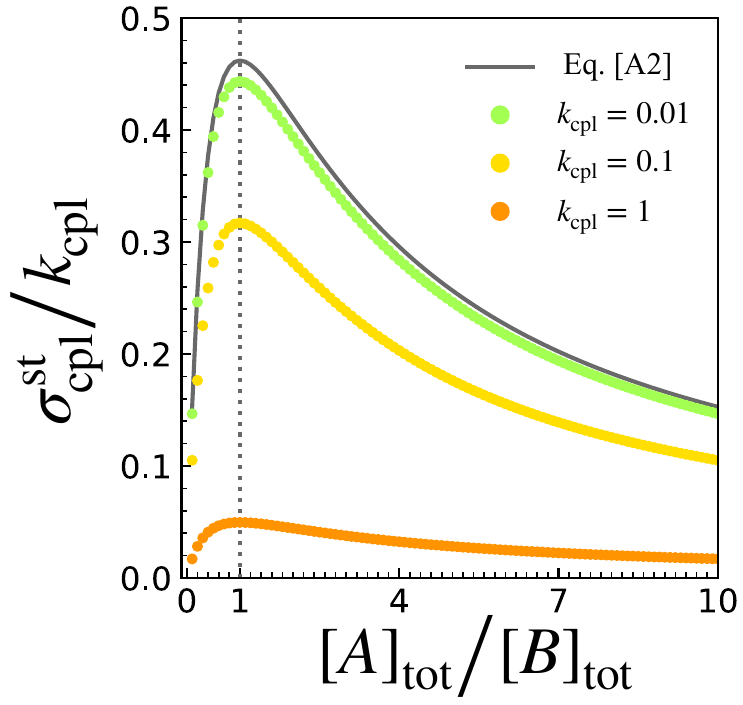}
    \caption{Dependence of $\sigma_\cpl^\st/\kc$ on the $\Atot/\Btot$ ratio with relatively weak currency coupling (colored points). The gray curve represents  {Eq.~}\eqref{eq:EPR_approx1}. 
        For a consistent comparison, the system size is kept constant by setting $\Ctot{\,=\,}\Atot{\,+\,}\Btot{\,=\,}2$ while varying $\Atot/\Btot$; parameters are set as $\kappa_A^+{\,=\,}2/3,\kappa_A^-{\,=\,}1/3$, $\kappa_B^+{\,=\,}1/3,\kappa_B^-{\,=\,}2/3$. 
    } \label{fig:EPR_approx1}
\end{figure}

Fig.~\ref{fig:EPR_approx1} numerically shows that the above approximation~\eqref{eq:EPR_approx1} holds well for small $\kc$. As $\kc$ increases, the (scaled) EPR $\sigma_\cpl^\st/\kc$ decreases overall and deviates from  {Eq.~}\eqref{eq:EPR_approx1}, but it is still maximized at $\Atot/\Btot{\,=\,}1$. 

\noindent\textbf{Strong currency coupling limit.} 
In the limit of strong currency coupling ($\kc\,{\gg}\,\kappa_X^\pm/\Xtot$), the charged/uncharged ratios at the steady state are approximated as:
\begin{align*}
    & \Gamma_A^\st = 
        \Gamma_0 + \frac{ \kappa_B \left( \kappa_A^+ - \Gamma_0\kappa_A^- \right) }{ \kc (\Atot\kappa_A^- + \Btot\kappa_B^-) }
        + O(k^{-2}_\cpl), \\
    & \Gamma_B^\st = 
        \Gamma_0 + \frac{ \kappa_A \left( \kappa_B^+ - \Gamma_0\kappa_B^- \right) }{ \kc(\Atot\kappa_A^- + \Btot\kappa_B^-) }
        + O(k^{-2}_\cpl).
\end{align*}
Accordingly, 
\begin{align*}
    & \Gamma_B^\st - \Gamma_A^\st \simeq 
    (\kappa_A^-\kappa_B^+ - \kappa_A^+\kappa_B^-)
        \frac{ 1 + \Gamma_0 }{ \kc(\Atot\kappa_A^- + \Btot\kappa_B^-) }, 
        \\
    & \frac{\Gamma_B^\st}{\Gamma_A^\st} \simeq 1 + 
    \frac{(\Atot\kappa_A + \Btot\kappa_B)\left( \kappa_A^-\kappa_B^+ - \kappa_A^+\kappa_B^- \right)}{\kc(\Atot\kappa_A^+ + \Btot\kappa_B^+)(\Atot\kappa_A^- + \Btot\kappa_B^-)}.
\end{align*}
With the definition of $\sigma^\st_\cpl$, this approximation leads to Eq.~\eqref{eq:EPR_approx_largek}: 
\begin{align} \label{eqS:EPR_approx_largek}
    \sigma^\st_\cpl \simeq\;& \frac{1}{\kc}
    \frac{ (\kappa_A^-\kappa_B^+ - \kappa_A^+\kappa_B^-)^2\Atot \Btot }{ (\Atot\kappa_A^+ + \Btot\kappa_B^+)(\Atot\kappa_A^- + \Btot\kappa_B^-) } 
    \nonumber\\
    \propto\;& \frac{1}{\kc}\frac{\frac{\Atot}{\Btot}}{ \left(\frac{\Atot}{\Btot}\kappa_A^+ + \kappa_B^+\right)\left(\frac{\Atot}{\Btot}\kappa_A^- + \kappa_B^-\right)}.
\end{align}

Moreover, from Eq.~\eqref{eqS:EPR_approx_largek} and Eq.~\eqref{eq:e_tilde_bounds}, $\sigma^\st_\cpl$ and $\tilde{e}_X$ are related as 
\begin{align} \label{eqS:tradeoff_k_large}
    & \sigma^\st_\cpl \simeq 
    \frac{(\kappa_A^-\kappa_B^+ - \kappa_A^+\kappa_B^-)^2}{\kc\sqrt{\kappa_A^+\kappa_A^- \kappa_B^+\kappa_B^- }} \sqrt{\tilde{e}^+_A\tilde{e}^-_A} \sqrt{\tilde{e}^+_B\tilde{e}^-_B} % \propto \tilde{e}_A\tilde{e}_B=\tilde{e}^2,
\end{align}
which leads to Eq.~\eqref{eq:tradeoff_k_large2}. 
This equation reveals a tradeoff: 
higher (lower) approximate elasticities are associated with larger (smaller) EPR. 
Given the constraint that the elasticities satisfy $\tilde{e}^\pm_{A}+\tilde{e}^\pm_{B}=1$, the geometric mean $\tilde{e}_A\tilde{e}_B$ of the elasticities becomes larger when the elasticities $\tilde{e}_X$ are evenly distributed. 

% {\color[cmyk]{0,.75,.90,0}
\noindent\textbf{Tradeoff inequality.} 
{We here} prove the inequality~\eqref{eq:tradeoff_inequality}. 

As $e=\left(e^+_{AA}e^-_{AA}e^+_{BB}e^-_{BB}\right)^{1/4}\geq\left(\tilde{e}^+_A\tilde{e}^-_A\tilde{e}^+_B\tilde{e}^-_B\right)^{1/4}$ holds from Eq.~\eqref{eq:e_tilde_bounds}, the left-hand side of {Eq.~}\eqref{eq:tradeoff_inequality} satisfies
\begin{align} \label{eqS:lhs_inequality}
    & \frac{(\kappa_A^-\kappa_B^+ - \kappa_A^+\kappa_B^-)^2}{\kc\sqrt{\kappa_A^+\kappa_A^- \kappa_B^+\kappa_B^- }} e^2\geq
    \frac{(\kappa_A^-\kappa_B^+ - \kappa_A^+\kappa_B^-)^2}{\kc\sqrt{\kappa_A^+\kappa_A^- \kappa_B^+\kappa_B^- }} \sqrt{\tilde{e}^+_A\tilde{e}^-_A} \sqrt{\tilde{e}^+_B\tilde{e}^-_B} 
    % \nonumber \\ & = 
    % \frac{(\kappa_A^-\kappa_B^+ - \kappa_A^+\kappa_B^-)^2}{\kc\sqrt{\kappa_A^+\kappa_A^- \kappa_B^+\kappa_B^- }} \frac{ \Atot\Btot\sqrt{\kappa_A^+\kappa_A^- \kappa_B^+\kappa_B^- } }
    % { (\Atot\kappa_A^+ + \Btot\kappa_B^+)( \Atot\kappa_A^- + \Btot\kappa_B^-)}
    \nonumber \\ & = 
    \frac{\Atot\Btot (\kappa_A^-\kappa_B^+ - \kappa_A^+\kappa_B^-)^2}
    { \kc (\Atot\kappa_A^+ + \Btot\kappa_B^+)( \Atot\kappa_A^- + \Btot\kappa_B^-)}.
\end{align}
Also, from $\log x \leq x-1$, the right-hand side of {Eq.~}\eqref{eq:tradeoff_inequality} satisfies
\begin{align} \label{eqS:rhs_inequality}
    \sigma^\st_\cpl & = 
    \kc\Atot\Btot\frac{\Gamma^\st_B - \Gamma^\st_A}{(\Gamma^\st_A+1)(\Gamma^\st_B+1)}\log\frac{\Gamma^\st_B}{\Gamma^\st_A}
    \nonumber \\ & \leq 
    \kc\Atot\Btot\frac{\Gamma^\st_B - \Gamma^\st_A}{(\Gamma^\st_A+1)(\Gamma^\st_B+1)}\left(\frac{\Gamma^\st_B}{\Gamma^\st_A}-1\right)
     \nonumber \\ & 
     = 
    \kc\Atot\Btot\frac{(\Gamma^\st_B - \Gamma^\st_A)^2}{\Gamma^\st_A(\Gamma^\st_A+1)(\Gamma^\st_B+1)} 
    .
\end{align}

% Therefore, it is sufficient to prove 
% \begin{align} \label{eqS:tradeoff_inequality2}
%     \frac{1}{\kc^2}\frac{ (\kappa_A^-\kappa_B^+ - \kappa_A^+\kappa_B^-)^2}
%     { (\Atot\kappa_A^+ + \Btot\kappa_B^+)( \Atot\kappa_A^- + \Btot\kappa_B^-)}
%     \geq \frac{(\Gamma^\st_B - \Gamma^\st_A)^2}{\Gamma^\st_A(\Gamma^\st_A+1)(\Gamma^\st_B+1)} .
% \end{align}
Noting that 
\begin{align} \label{eqS:rhs_inequality}
    \Gamma^\st_B - \Gamma^\st_A = & 
    \frac{
        (\kappa_A^-\kappa_B^+ - \kappa_A^+\kappa_B^-)
    }{
        (\kc\Atot\kappa_A^- + \kc\Btot\kappa_B^- + \kappa_A^-\kappa_B)
    }
    \nonumber \\ & \times
    \frac{
        (\kc\Atot\kappa_A + \kc\Btot\kappa_B + \kappa_A\kappa_B)
    }{
        (\kc\Atot\kappa_A^- + \kc\Btot\kappa_B^- + \kappa_A\kappa_B^-)
    },
\end{align} 
it is sufficient to prove
\begin{align} \label{eqS:tradeoff_inequality2}
    & \frac{1}
    { \kc^2(\Atot\kappa_A^+ + \Btot\kappa_B^+)( \Atot\kappa_A^- + \Btot\kappa_B^-)}
    \nonumber \\ & \geq 
    \frac{1}{
        (\kc\Atot\kappa_A^+ + \kc\Btot\kappa_B^+ + \kappa_A^+\kappa_B)
    }
    \nonumber \\ & \quad \times
    \frac{1}{
        (\kc\Atot\kappa_A^- + \kc\Btot\kappa_B^- + \kappa_A\kappa_B^-)
    }.
\end{align}
As {Eq.~}\eqref{eqS:tradeoff_inequality2} is evident, we obtain Eq.~\eqref{eq:tradeoff_inequality}.
% }

\subsection{Bounds on $\Gamma_X$ modulation range} \label{sec:bound_modulation_range}
\begin{figure}[tb]
    \centering \includegraphics[width=0.98\linewidth, clip]{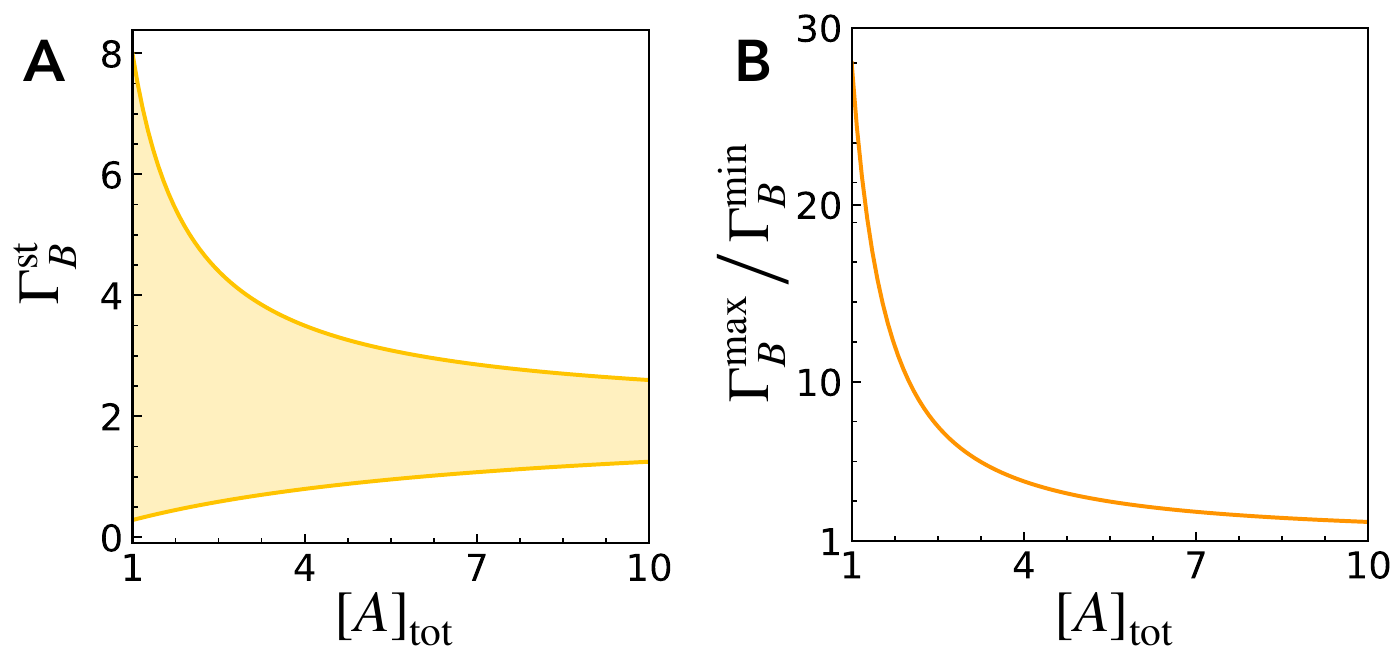} 
    \caption{
        Dependence of (A) $\Gamma_B^\st$ modulation range, $[\Gamma_B^{\min},\Gamma_B^{\max}]$, and (B) its width, $\Gamma_B^{\max}/\Gamma_B^{\min}$, 
        on $\Atot$. $\kappa_B{\,=\,}1$, $\kc{\,=\,}1$, and $\Btot{\,=\,}1$ are fixed. 
    } \label{fig:overall_control}
\end{figure}

In addition to controllability in terms of elasticity, the achievable ranges of the charged/uncharged ratios $\Gamma_X^\st$ are fundamentally constrained by {finite driving strength, $\kappa_X^\pm$. If the total driving strength $\kappa_X{\,=\,}\kappa_X^++\kappa_X^-$} has an upper bound $\kappa_X^{\max}$, $\Gamma_X^\st$ is bounded as $\Gamma_X^{\max}{\,\geq\,}\Gamma_X^\st{\,\geq\,}\Gamma_X^{\min}$ with 
\begin{eqnarray*}
    \Gamma_{X}^{\max} &=& 
    \frac{\kappa_Y^+}{\kappa_Y^-} + \frac{ \kc\Xtot + \kappa_Y }{ \kc\Ytot\kappa_Y^- }\kappa_X^{\max}, % \label{eq:B_ratio_max} 
    \\
    \Gamma_{X}^{\min} &=& 
    \left( \frac{\kappa_Y^-}{\kappa_Y^+} + \frac{ \kc\Xtot + \kappa_Y }{ \kc\Ytot\kappa_Y^+ } \kappa_X^{\max} \right)^{-1}.
    \quad(Y\,{\neq}\,X)  
    % \label{eq:B_ratio_min} 
\end{eqnarray*}
These upper and lower bounds are governed by the relative magnitudes of $\Xtot\kappa_X^\pm$ and $\Ytot\kappa_Y^\pm$. When the currency metabolite pools are imbalanced ($\Atot\,{\gg}\,\Btot$), the charged/uncharged ratio $\Gamma_B^\st$ of the less abundant currency metabolite $B$ is confined to a narrower range $\Gamma_{B}^{\max} / \Gamma_{B}^{\min}$ (Fig.~\ref{fig:overall_control}), matching the trend observed in the elasticities. 
    % An increase in the bias of $\Xtot$ toward $\Atot$ reduces the attainable range of $\Gamma_{B}^\st$, $\Gamma_{B}^{\max} / \Gamma_{B}^{\min}$, which is consistent with the trend in elasticity . 

Therefore, to achieve the wide range of charged/uncharged ratios observed \textit{in vivo}, cells must regulate not only the bias $\kappa_X^+ / \kappa_X^-$ of the driving reactions for currency metabolite $X$ but also the pool sizes $\Xtot$ of currency metabolites.

% \newpage
\section{On the correlation between the intracellular ATP/GTP ratio and GC content}\label{sec:GTP-GC}
The intracellular concentration $[{\rm NTP}]$ of NTPs (i.e., ATP and GTP) is determined by the balance between their production flux $P_{\rm NTP}$ and their conversion and dilution:
\begin{eqnarray*}
    \frac{{\rm d}}{{\rm d}t}[{\rm NTP}] = P_{\rm NTP} - k_{\rm NTP}[{\rm NTP}].
\end{eqnarray*}
Then, their steady-state concentration is
\begin{eqnarray*}
    [{\rm NTP}]^\st = \frac{P_{\rm NTP}}{k_{\rm NTP}}. 
\end{eqnarray*}

One might reasonably expect the production flux of GTP and guanine to increase with genomic GC content, as supported by experimental evidence~\cite{dietel2019selective}. 
We therefore assume that the production fluxes of ATP and GTP, $P_{\rm ATP}$ and $P_{\rm GTP}$, positively correlate with the genomic AT and GC contents, respectively. 

{Furthermore, if $P_{\rm NTP}$ is decomposed into the flux used for RNA incorporation and that for \textit{de novo} synthesis for currency metabolites, it is natural to assume that only the former is proportional to AT or GC content. Nevertheless, because the pool of free NTPs derived from RNA turnover can greatly exceed that from \textit{de novo} synthesis, $P_{\rm ATP}$ and $P_{\rm GTP}$ are still expected to correlate positively with AT and GC content, respectively.} 
    According to ref.~\cite{milo2015cell}, the weight of RNA per cell is about $60$ fg, and thus that of ATP or GTP is about $15$ fg. In contrast, the intracellular concentrations of ATP and GTP are about $3$ mM and $1$ mM, respectively~\cite{danchin1984metabolic,jewett2009continued,deng2021measuring}; because the molecular weights of ATP and GTP are $507$ and $523$ $\rm g/mol$ and the volume of microbes such as \textit{E. coli} is about $1\,{\rm \mu m^3}$, % $= 10^{-18}\;{\rm m}^3$, 
    the weights of (free) ATP and GTP are about $1.5$ fg and $0.5$ fg, respectively. That is, the intracellular abundance of guanine molecules for RNA is higher by more than an order of magnitude. 

{In contrast to the above causal direction, in which GC content influences NTP pool sizes, it has also been suggested that the GTP pool size~\cite{mrnjavac2025gtp} as well as the abundance of resources required for DNA/RNA synthesis, such as nitrogen sources~\cite{seward2016dietary}, can bias genome GC content. In any case, some correlation between NTP pool sizes as currency metabolites and genomic GC content is expected.}

% {\color[cmyk]{0,.75,.90,0}
{\color{black}
\section{Robustness of the cost–controllability tradeoff in generalized currency coupling} \label{sec:GeneralizedCoupling}

% {\color[cmyk]{0,.75,.90,0}
% {\color[cmyk]{1,.45,0,0}
\subsection{Ping-pong bi-bi mechanism in currency coupling} \label{sec:PingPong}
In the main text, we assumed the mass action kinetics for the coupling reaction: i.e., $J_\cpl=\kc([A][B^\ast] - [A^\ast][B])$. In other words, an ideal dilute solution was assumed for simplicity. 

In general, this assumption is not necessarily realistic. For example, nucleoside diphosphate kinase (NDPK) catalyzes the coupling of NTPs via the so-called ping-pong bi-bi mechanism~\cite{gonin1999catalytic}. In this case, a more realistic functional form for $J_\cpl$ is represented as~\cite{liebermeister2006bringing,imperial2014enzyme}:
\begin{align}
    & J_{\mathrm{cpl}}
    = k_{\mathrm{cpl}}([A][B^\ast] - [A^\ast][B]) \Big/ \Bigl({\footnotesize 
    \frac{[A]}{K_{A}} + \frac{[B^\ast]}{K_{B^\ast}} 
    + \frac{[A][B^\ast]}{K_{A} K_{B^\ast}} }
    \nonumber \\ 
    & {\footnotesize + \frac{[A^\ast]}{K_{A^\ast}} + \frac{[B]}{K_{B}} 
    + \frac{[A^\ast][B]}{K_{A^\ast} K_{B}} 
    + \frac{[A][A^\ast]}{K_{A} K_{A^\ast}} 
    + \frac{[B^\ast][B]}{K_{B^\ast} K_{B}} 
    }\Bigr), 
    \label{eq:PingPong}
\end{align}
with parameters $K_X$ and $K_{X^\ast}$ ($X=A,B$). 

% \gray{Likewise, in the case of a general Mikaelis-Menten kinetics, ...~\cite{liebermeister2006bringing}.} 

Even with this reaction kinetics, the results in the main text remain qualitatively unchanged, and a cost–controllability tradeoff is numerically yielded (Fig.~\ref{fig:PingPong}). 
% }

\begin{figure*}[tb]
    \centering 
    \includegraphics[width=\linewidth, clip]{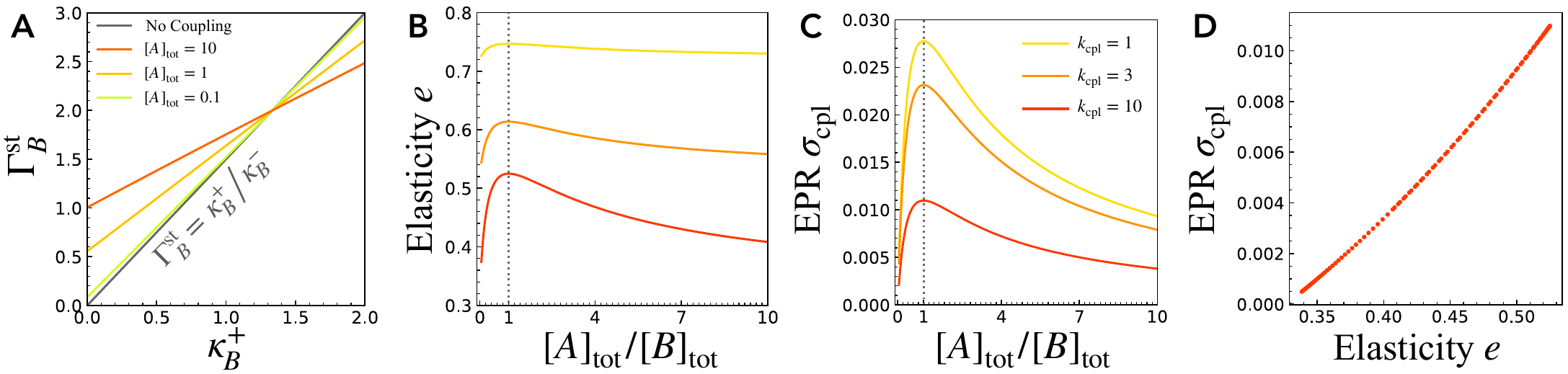} 
    \caption{{
    Cost–controllability tradeoff in metabolic currency coupling with a ping-pong bi-bi mechanism: $
    J_{\mathrm{cpl}}
    =k_{\mathrm{cpl}}
    \frac{
    \,[A][B^\ast] - \,[A^\ast][B]
    }{
    [A] + [B^\ast] + [A][B^\ast]
    + [A^\ast] + [B] + [A^\ast][B] 
    + [A][A^\ast] + [B^\ast][B] 
    }$, where parameters in Eq.~\eqref{eq:PingPong}, $K_X$ and $K_{X^\ast}$, are set as unity. 
    (A) Dependence of $\Gamma_B^\st$ on $\kappa_B^+$. $\Btot=1$, $\kc=10$. 
    (B) Dependence of mean elasticity $e$ on $\Atot/\Btot$. 
    (C) Dependence of EPR $\sigma^\st_\cpl$ on $\Atot/\Btot$. 
    (D) Relationship between mean elasticity $e$ and EPR $\sigma^\st_\cpl$. $\kc=10$. 
    In (B-D), $\Atot+\Btot=1$ and $\kappa_B^+{\,=\,}1/3$ are fixed. 
    In (A-D), the other parameters are set as $\kappa_A^+{\,=\,}2/3$, $\kappa_A^-{\,=\,}1/3$, and $\kappa_B^-{\,=\,}2/3$. 
    }} \label{fig:PingPong}
\end{figure*}

\subsection{Stoichiometry mismatch and dissipation} 
% \label{sec:GeneralizedCoupling}
In the main text, for simplicity, we assumed $\Delta\mu_A^\circ=\Delta\mu_B^\circ$, so that $k_{\mathrm{cpl}}^{+}=k_{\mathrm{cpl}}^{-}$, and we considered a $1{:}1$ coupling reaction in the form of $A+B^\ast \rightleftharpoons A^\ast+B$.
These assumptions are reasonable for coupling within the same group of currencies, such as among NTPs or between NADPH and NADH. However, they do not necessarily hold for other currency couplings, including the electron transport chain (ETC), where the ATP/ADP and NADH/NAD couples are linked, and coupling between ATP/ADP and ADP/AMP. 

Nevertheless, our main conclusions remain essentially unchanged for the more general case with $\Delta\mu_A^\circ\neq\Delta\mu_B^\circ$ and a stoichiometric exchange $\nu A+B^\ast \rightleftharpoons \nu A^\ast+B$. 
% \noindent\textbf{Case without dissipation in the coupling reaction.}
In this case, assuming the local detailed balance condition, the coupling rate constants satisfy
\begin{equation}
    k_{\mathrm{cpl}}^{+}
    =
    k_{\mathrm{cpl}}^{-}\exp\!\left(\frac{\Delta\mu_B^\circ-\nu\Delta\mu_A^\circ}{RT}\right),
    \label{eq:k_ratio_general}
\end{equation}
where $R$ is the gas constant and $T$ is the temperature. The dynamics can then be written as {\small 
\begin{eqnarray*}
    \frac{{\rm d}}{{\rm d}t}[A] \!&=&\!
    -\nu\!\left(k_{\mathrm{cpl}}^{+}[A][B^\ast]-k_{\mathrm{cpl}}^{-}[A^\ast][B]\right)
        -\kappa_A^{+}[A]+\kappa_A^{-}[A^\ast], \nonumber\\
    \frac{{\rm d}}{{\rm d}t}[A^\ast] \!&=&\!
    \nu\!\left(k_{\mathrm{cpl}}^{+}[A][B^\ast]-k_{\mathrm{cpl}}^{-}[A^\ast][B]\right)
        +\kappa_A^{+}[A] - \kappa_A^{-}[A^\ast], \nonumber\\
    \frac{{\rm d}}{{\rm d}t}[B] \!&=&\!
    \left(k_{\mathrm{cpl}}^{+}[A][B^\ast]-k_{\mathrm{cpl}}^{-}[A^\ast][B]\right)
    -\kappa_B^{+}[B]+\kappa_B^{-}[B^\ast], \nonumber\\
    \frac{{\rm d}}{{\rm d}t}[B^\ast] \!&=&\!
    -\left(k_{\mathrm{cpl}}^{+}[A][B^\ast]-k_{\mathrm{cpl}}^{-}[A^\ast][B]\right)
        +\kappa_B^{+}[B] - \kappa_B^{-}[B^\ast].
    \label{eq:dyn_general_nu}
\end{eqnarray*}
}At steady state, this stoichiometry effectively rescales the driving reactions for currency metabolite $A$ as $\kappa_A^{\pm}\to \kappa_A^{\pm}/\nu$.
Aside from this rescaling and $k_{\mathrm{cpl}}^{+}\neq k_{\mathrm{cpl}}^{-}$, the results in the main text remain unchanged.

\noindent\textbf{Case with free-energy dissipation in the coupling reaction.} 
In ETC-like settings, the proton-motive force is dissipated, and consequently the effective coupling between redox currencies (e.g., NADH/NAD) and phosphorylation currencies (ATP/ADP) can vary, even for the same amount of NADH oxidized. In such situations, the coupling step cannot be treated as microscopically reversible in isolation, and the local-detailed-balance condition Eq.~\eqref{eq:k_ratio_general} need not apply directly at the coarse-grained level. 

To incorporate such dissipation, we can additionally consider ``leakage reactions'' that effectively decrease the conversion stoichiometry in the coupling reactions between $A$ and $B$: 
\begin{align}
    & A + B^\ast + C \overset{k_{\rm leak}}{\to} A + B + C^\ast, \nonumber\\
    & A^\ast + B + C \overset{k_{\rm leak}}{\to} A + B + C^\ast,
    \label{reac:dissipation}
\end{align}
where we assume the concentration of $C$ (and $C^\ast$) is chemostated for simplicity and the effective rate constants are then denoted as $k_{\rm leak}$. 
Then, by assuming mass action kinetics, $d:=k_{\rm leak}[C]/(k_{\rm leak}[C]+\kc)$ ($0< d<1$) parametrizes the degree of dissipation in the coupling reaction $A+B^\ast \rightleftharpoons A^\ast+B$. The resulting dynamics are{\small 
\begin{eqnarray*}
    \frac{{\rm d}}{{\rm d}t}[A] \!&=&\! 
    - k_{\mathrm{cpl}}[A][B^\ast] + \frac{1}{1-d}k_{\mathrm{cpl}}[A^\ast][B] 
    -\kappa_A^{+}[A]+\kappa_A^{-}[A^\ast], \nonumber\\
    \frac{{\rm d}}{{\rm d}t}[A^\ast] \!&=&\! 
    k_{\mathrm{cpl}}[A][B^\ast] - \frac{1}{1-d}k_{\mathrm{cpl}}[A^\ast][B] 
    +\kappa_A^{+}[A]-\kappa_A^{-}[A^\ast], \nonumber\\
    \frac{{\rm d}}{{\rm d}t}[B] \!&=&\! 
    \frac{1}{1-d}k_{\mathrm{cpl}}[A][B^\ast] - k_{\mathrm{cpl}}[A^\ast][B] 
    -\kappa_B^{+}[B]+\kappa_B^{-}[B^\ast], \nonumber\\
    \frac{{\rm d}}{{\rm d}t}[B^\ast] \!&=&\! 
    - \frac{1}{1-d}k_{\mathrm{cpl}}[A][B^\ast] + k_{\mathrm{cpl}}[A^\ast][B] 
    +\kappa_B^{+}[B]-\kappa_B^{-}[B^\ast].
    \label{eq:dyn_dissipative}
\end{eqnarray*}
}Accordingly, the coupling reaction consumes $\frac{1}{1-d}\,({>}1)$ units of $A^\ast$ to obtain $1$ unit of $B^\ast$, and \textit{vice versa}. Even in this dissipative case, as long as $d$ is sufficiently small, the cost–controllability tradeoff qualitatively holds (Fig.~\ref{figSI:tradeoff_disspation}).
% \\ cf. loose coupling by Oosawa?
}

\begin{figure}[bt]
    \centering 
    \includegraphics[width=0.8\linewidth, clip]{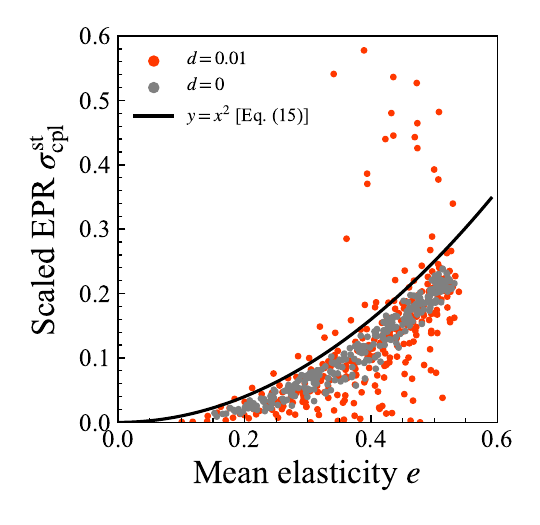}
    \caption{{Thermodynamic cost–controllability tradeoff with dissipation in currency coupling. 
        The scaled EPR of the coupling reaction, $\frac{\kc \sqrt{\kappa_A^+\kappa_A^- \kappa_B^+\kappa_B^-} }{(\kappa_A^-\kappa_B^+ - \kappa_A^+\kappa_B^-)^2} \sigma^\st_\cpl$, is plotted against mean elasticity $e$ for the dissipative case (orange; $d=0.01$) and the non-dissipative case (gray; $d=0$; see also Fig.~\ref{fig:EPR}B). The parameters $\kappa_X^\pm$ and $\Xtot$ are randomly sampled from a uniform distribution within the range $[0,1]$. $\kc{\,=\,}10$. 
        The black curve represents the strong coupling limit without dissipation, Eq.~\eqref{eq:tradeoff_k_large2}.} 
    } \label{figSI:tradeoff_disspation}
\end{figure}

\section{Order estimation of model parameter values}\label{sec:order}
It is thought that cells generally maintain substantial levels of enzymes responsible for metabolic currency coupling. NDPK, for example, is considered a housekeeping enzyme that is constitutively expressed in many types of cells to maintain nucleotide homeostasis~\cite{yu2017nucleoside}.
    % (\url{https://www.perplexity.ai/search/rate-constant-for-the-nucleosi-uw71B64TQ42H7yzgB4j7lw}). 

To evaluate the biological relevance of our theory, particularly the controllability–EPR tradeoff relationship, we estimate typical values of relevant parameters and variables, using ATP–GTP coupling as an example (summarized in Table~\ref{table:order}). 

\noindent\textbf{Effective rate constant of driving reaction.} 
    % \url{https://elifesciences.org/reviewed-preprints/98800} などで使われている力学系モデルの数値計算における値: 
    % \url{https://www.embopress.org/doi/full/10.15252/msb.202110504} \url{https://github.com/yhimeoka/Perturbation-Response-Analysis}  \url{https://www.sciencedirect.com/science/article/pii/S1096717614000731#f0025}
    % \\- GTPはモデルにないが, $\kappa_X$のorder estimationには使える
The total ATP synthesis flux in a microbe is estimated as~\cite{deng2021measuring}
\begin{eqnarray*}
    \kappa^+_{\rm ATP}\ADP \simeq 
    10^{-2}\;{\rm [M/s]},
\end{eqnarray*}
and the intracellular ADP concentration in \textit{E. coli} is measured as~\cite{albe1990cellular} 
\begin{eqnarray*}
    \ADP \simeq 10^{-3}\;{\rm [M]}. 
\end{eqnarray*}

Therefore, the effective rate constant for the driving reaction, ${\rm ADP \leftrightarrow ATP}$, should be estimated as 
\begin{eqnarray*}
    \kappa^+_{\rm ATP} \simeq 
    10^1 \;{\rm [s^{-1}]}.
\end{eqnarray*}

{Given that $\ATP_\tot{\,\simeq\,} 3\times10^{-3}\;{\rm [M]}~\cite{albe1990cellular}$}, 
\begin{eqnarray} \label{eq:kappa_est}
    \kappa^+_{\rm ATP}/\ATP_\tot
    \simeq 10^3\text{--}10^4 \;{\rm [M^{-1}s^{-1}]}.
\end{eqnarray}

\noindent\textbf{Coupling constant $\kc$ in ping-pong mechanism.}
% The typical value of the coupling constant $\kc$ is here estimated,  in the case of ATP–GTP coupling via NDPK. 
NDPK catalyzes the following reactions:
\begin{eqnarray} \label{reac:NDPK}
    {\rm NTP} + {\rm E} 
    \overset{k^+_N}{\underset{k^-_N}{\rightleftharpoons}} 
    {\rm NDP} + {\rm EP}
    \quad{\rm (N=A,G)}, 
\end{eqnarray}
in the case of ATP–GTP coupling. 
Through this ping-pong mechanism, the charged/uncharged ratios, $\Gamma_{\rm ATP}$ and $\Gamma_{\rm GTP}$, are driven toward equilibrium, providing an effective coupling between them. 

Assuming that reaction~\eqref{reac:NDPK} follows mass action kinetics, the steady-state concentrations of E and EP are calculated as
\begin{eqnarray*}
    {\rm [E]}^\st &=& \frac{k_A\ADP + k_G\GDP}{k_A\ATP_\tot + k_G\GTP_\tot}\Etot, \\
    {\rm [EP]}^\st &=& \frac{k_A\ATP + k_G\GTP}{k_A\ATP_\tot + k_G\GTP_\tot}\Etot,
\end{eqnarray*}
where $\Etot$ denotes the total amount of NDPK, and we assume $k_N{\,:=\,}k^+_N{\,=\,}k^-_N$. 

Accordingly, we obtain the following dynamics of ${\ATP}$: 
\begin{align*}
    \frac{d}{dt}{\ATP} =\;& k_A\ADP{\rm [EP]}^\st - k_A\ATP{\rm [E]}^\st 
    \\\;& + \kappa_A^+\ADP - \kappa_A^-\ATP  \\
    =\;& k_A\Etot\frac{ k_G\ADP\GTP - k_G\ATP\GDP}{k_A\ATP_\tot + k_G\GTP_\tot}
    \\\;& + \kappa_A^+\ADP - \kappa_A^-\ATP.
\end{align*}
It follows 
\begin{eqnarray} \label{eq:kc_mass_action}
    \kc = \frac{ k_Ak_G\Etot}{k_A\ATP_\tot + k_G\GTP_\tot}.
\end{eqnarray}

From refs.~\cite{zala2017advantage,tokarska2008nucleoside,pollack2002suspected}, 
\begin{eqnarray*}
    k_{\rm cat} &\simeq& 10^3 \;[{\rm /s}]
    % 0.7\sim13\times10^6 \;[{\rm M^{-1}s^{-1}}]
    , \\ 
    K_{\rm M, ATP} &\simeq& 10^{-4}\text{--}10^{-3} \;[{\rm M}]
    , \\ 
    K_{\rm M, GDP} &\simeq& 10^{-4} \;[{\rm M}]
    , \\ 
    \Etot &\simeq& 10^{-6} \;[{\rm M}],
\end{eqnarray*}
and thus, 
\begin{eqnarray}
    k_N \simeq k_{\rm cat} / K_{\rm M,N} \simeq 10^6\text{--}10^7 {\rm [M^{-1}s^{-1}]}. 
\end{eqnarray}
This is consistent with ref.~\cite{schaertl1998substrate}, which states that the second order rate constants $k_N$ for phosphorylation by natural NTPs are between $0.7\text{--}13\times10^6 \;{\rm [M^{-1}s^{-1}]}$. 

Finally, from {Eq.~}\eqref{eq:kc_mass_action}, 
\begin{eqnarray} \label{eq:kc_est}
    \kc \simeq 10^3\text{--}10^4 {\rm [M^{-1}s^{-1}]}.
\end{eqnarray}

% \vspace{10pt}
From Eqs.~\eqref{eq:kappa_est}~and~\eqref{eq:kc_est}, $\kc$ and $\kappa^+_{\rm ATP}/\ATP_\tot$ are typically in the same order of magnitude, while they depend largely on the amount of enzymes and the growth conditions. 
This order-of-magnitude estimation suggests that $\kc$ and $\kappa_{\rm ATP}^\pm/[{\rm ATP}]_\tot$ are of comparable magnitude with typical NDPK concentrations, implying that both strong ($\kc\,{\gg}\,\kappa_{\rm ATP}^\pm/[{\rm ATP}]_\tot$) and weak ($\kc\,{\ll}\,\kappa_{\rm ATP}^\pm/[{\rm ATP}]_\tot$) currency coupling may occur in natural microbial systems. 
In particular, under nutrient-limited conditions, the strong coupling limit will be more likely.

\section{Coupling of $n\geq3$ Currency Metabolites}\label{sec:n=3}
As an extension of the minimal model discussed in the main text, we consider a system with $n\geq3$ kinds of currency metabolites, $X_i\;(i=1,2,\cdots,n)$. 
Each metabolite $X_i$ is driven by a reaction with effective rate constants $\kappa^\pm_i$: 
\begin{eqnarray*}
    X_i 
    \underset{\kappa^+_i}{\overset{\kappa^-_i}{{\rightleftharpoons}}} 
    X_i^\ast \quad (i=1,2,\cdots,n).
\end{eqnarray*}

These currency metabolites are also assumed to be coupled through exchange reactions of the form\footnote{The following analysis can be applied to systems in which the coupling constant varies among currency metabolite pairs.}:
\begin{eqnarray*}\label{SI_reac:exchange}
    X_i + X_j^\ast 
    \underset{\kc}{\overset{\kc}{{\rightleftharpoons}}} 
    X_i^\ast + X_j.
    % \quad(i<j)
\end{eqnarray*}

Assuming mass-action kinetics, the rate equation is given by  
\begin{align}
    \dot{[X^\ast_i]} =\; & \kappa^+_i[X_i] - \kappa^-_i[X^\ast_i] + \sum_{i\neq j}\kc([X_i][X^\ast_j] - [X^\ast_i][X_j]) \nonumber \\
    = \; & \kappa^+_i[X_i]_\tot - \kappa_i[X^\ast_i]  
    \nonumber \\ & + 
    \sum_{j \neq i}\kc( [X_i]_\tot[X^\ast_j] - [X^\ast_i][X_j]_\tot ), \label{eq:n3_dynamics}
\end{align}
where $\kappa_i:=\kappa^+_i+\kappa^-_i$ $(i=1,2,\cdots,n)$.

\noindent\textbf{Steady state.} From {Eq.~}\eqref{eq:n3_dynamics}, the steady state $[X^\ast_i]^\st$ satisfies
\begin{eqnarray*} 
    \left( \D+\R \right){\bf [X^\ast]^\st} = \{ \kappa_i^+[X_i]_\tot \}_i, 
\end{eqnarray*}
where a diagonal matrix ${\bf D}$ and a rank-one matrix ${\bf R}$ are defined as 
\begin{eqnarray*}
    {\bf D} := {\rm diag}(\kappa_i+\kc \Ctot)_i, \quad
    {\bf R} := -\{ \kc[X_i]_\tot \}_i\cdot{\bf 1}^\top, 
\end{eqnarray*}
with $\Ctot{\,:=\,}\sum_{j}[X_j]_\tot$.

Therefore, we have: 
\begin{eqnarray*}
        {\bf [X^\ast]}^\st = (\D+\R)^{-1}\{ \kappa_i^+[X_i]_\tot \}_i . 
\end{eqnarray*}
Here, the inverse can be evaluated using the matrix inversion lemma (Woodbury matrix identity):
\begin{align*} 
    & (\D+\R)^{-1} = \D^{-1} + 
    \\& \;\; \D^{-1} \{\kc[X_i]_\tot \}_i (1 - {\bf 1}^\top\D^{-1}\{\kc[X_i]_\tot \}_i)^{-1} {\bf 1}^\top\D^{-1}.
\end{align*}
Accordingly, we obtain:
{\small 
\begin{align*}
    & [X^\ast_i]^\st = \frac{[X_i]_\tot}{\kappa_i +\kc \Ctot} \left(
        \kappa_i^+ + \kc
        \frac{\sum_j  \frac{\kappa_j^+ [X_j]_\tot}{\kappa_j +\kc \Ctot} }
        { 1 - 
        \sum_j  \frac{\kc [X_j]_\tot}{\kappa_j +\kc \Ctot} }
    \right), \\
    & [X_i]^\st = \frac{[X_i]_\tot}{\kappa_i +\kc \Ctot} \left(
        \kappa_i^- + \kc
        \frac{\sum_j  \frac{\kappa_j^- [X_j]_\tot}{\kappa_j +\kc \Ctot} }
        { 1 - 
        \sum_j  \frac{\kc [X_j]_\tot}{\kappa_j +\kc \Ctot} }
    \right), 
\end{align*}
}
\begin{eqnarray} \label{eq:SI2_ratio}
    \Gamma_i^\st :=\frac{[X^\ast_i]^\st}{[X_i]^\st}
    = \frac{
        \kappa_i^+ + 
        \kc\sum_{j \neq i}  \frac{(\kappa_j^+ - \kappa_i^+) [X_j]_\tot}{\kappa_j +\kc \Ctot}
    }{
        \kappa_i^- + 
        \kc\sum_{j \neq i}  \frac{(\kappa_j^- - \kappa_i^-) [X_j]_\tot}{\kappa_j +\kc \Ctot}
    }. 
\end{eqnarray}

In the strong coupling limit, 
\begin{align}
    \Gamma_i^\st = &\; \Gamma_0 + k_\cpl^{-1}
    \frac{
        \sum_{j \neq i} \kappa_j[X_j]_\tot 
            \left( ( \kappa_i^+ - \kappa_j^+ ) - \Gamma_0( \kappa_i^- - \kappa_j^- ) \right)
        }{\Ctot \sum_{j} [X_j]_\tot \kappa_j^- }
    \nonumber \\ 
    &  + O(k_\cpl^{-2}), \label{eq:SI_n_Ratio_StrongLimit}
\end{align}
with 
\begin{align*} 
    \Gamma_0 := \frac{
        \sum_{j} [X_j]_\tot \kappa_j^+ 
    }{
        \sum_{j} [X_j]_\tot \kappa_j^- 
    }.
\end{align*}

% These results are consistent with the case of $n=2$ in the main text (Eqs.~(4)-(6)).

% \vspace{10pt}
\noindent\textbf{Elasticity.} 
The dimensionless elasticity is defined as
\begin{align*}
    e^\pm_{ij} := \pm\frac{\partial \log\Gamma_i^\st}{\partial \log\kappa^\pm_j}.
\end{align*}
Then, 
\begin{align*}
    & e^\pm_{ii} 
    = 1 - \frac{ 
        \kc\sum_{l \neq i}  \frac{\kappa_l^\pm[X_l]_\tot}{\kappa_l +\kc \Ctot}
    }{
        \kappa_i^\pm + 
        \kc\sum_{l \neq i}  \frac{(\kappa_l^\pm - \kappa_i^\pm) [X_l]_\tot}{\kappa_l +\kc \Ctot}
    }
    , \\& 
    e^\pm_{ij} = \frac{ 
        \kc \frac{\kappa^\pm_j [X_j]_\tot}{\kappa_j +\kc \Ctot} 
    }{
        \kappa_i^\pm + 
        \kc\sum_{l \neq i}  \frac{(\kappa_l^\pm - \kappa_i^\pm) [X_l]_\tot}{\kappa_l +\kc \Ctot}
    } 
    \;\;(j \neq i). 
\end{align*}
Accordingly, $\sum_j e^\pm_{ij} = 1$. 

In the strong coupling limit, 
\begin{align*}
    e^\pm_{ij} \to \tilde{e}_l := \frac{ 
        [X_j]_\tot \kappa^\pm_j 
    }{  \sum_{l} [X_l]_\tot \kappa_l^\pm 
    }. 
\end{align*}

\noindent\textbf{EPR and tradeoff.} 
The EPR of all coupling reactions is given by
\begin{align*}
        \sigma_\cpl^{(n)} :=\;& \sum_{i>j} \sigma_{(i,j)}^\st
        \\ =\;& 
        \sum_{i>j}\kc [X_i]_\tot[X_j]_\tot 
        \frac{\Gamma_i^\st-\Gamma_j^\st}{(\Gamma_i^\st+1)(\Gamma_j^\st+1)}
        \log\frac{\Gamma_i^\st}{\Gamma_j^\st}. 
\end{align*}

From {Eq.~}\eqref{eq:SI_n_Ratio_StrongLimit}, in the strong coupling limit, each $\sigma_{(i,j)}^\st$ can be approximated as: 
{\small
\begin{align*}
        \sigma_{(i,j)}^\st & \simeq 
             \frac{k_\cpl^{-1} [X_i]_\tot [X_j]_\tot}{(1+\Gamma_0)^2 \Gamma_0}
            \left( \frac{d\Gamma_i^\st}{d k_\cpl^{-1}}\big|_{k_\cpl^{-1}\to 0} - \frac{d\Gamma_j^\st}{d k_\cpl^{-1}}\big|_{k_\cpl^{-1}\to 0} \right)^2
        % \\ &= \frac{k_\cpl^{-1} [X_i]_\tot [X_j]_\tot}{(1+\Gamma_0)^2\Gamma_0}
        % \left( 
        % \frac{
        %     \sum_l \kappa_l[X_l]_\tot 
        %     \left( ( \kappa_i^+ - \Gamma_0\kappa_i^+ ) - ( \kappa_j^+ - \Gamma_0\kappa_j^+ ) \right)
        % }{\Ctot \sum_l [X_l]_\tot \kappa_l^- } \right)^2
        \\ & = \frac{[X_i]_\tot [X_j]_\tot}{k_\cpl(\sum_l [X_l]_\tot \kappa_l^+)(\sum_l [X_l]_\tot \kappa_l^-)}
        \times \\ & \quad
        \left(
            \frac{\sum_{l}[X_l]_\tot\left( 
                (\kappa_i^+-\kappa_j^+)\kappa_l^- - (\kappa_i^--\kappa_j^-)\kappa_l^+
            \right)}{\Ctot}
        \right)^2.
\end{align*}
}That is,
{\small
\begin{align}
        & \sigma_{(i,j)}^\st \simeq  
        \frac{\left(
                (\kappa_i^+-\kappa_j^+)\overline{\kappa^-} - (\kappa_i^--\kappa_j^-)\overline{\kappa^+}
        \right)^2}{\kc\sqrt{\kappa_i^+\kappa_i^-}\sqrt{\kappa_j^+\kappa_j^-}}
        % \times \nonumber \\ & \quad \quad \quad \;\;  
        \sqrt{\tilde{e}^+_i\tilde{e}^-_i} \sqrt{\tilde{e}^+_j\tilde{e}^-_j}
        , \label{eq:SI_n_EPR_StrongLimit}
\end{align}
}where $\overline{\kappa^\pm}:=(\sum_l[X_l]_\tot\kappa_l^\pm)/(\sum_l[X_l]_\tot)$ is the averaged effective rate constant of all driving reactions. 
Therefore, a tradeoff between the EPR $\sigma_\cpl^{(n)}$ and elasticity in the strong coupling limit (analogous to Eq.~\eqref{eq:tradeoff_k_large2}) still holds, although it is biased by the driving reactions for currency metabolites other than $X_i$ and $X_j$.

Noting the constraint $\sum_i \tilde{e}^\pm_i=1$, {Eq.~}\eqref{eq:SI_n_EPR_StrongLimit} implies that the more uniformly large the values of $\tilde{e}^\pm_i$ are, the greater the total EPR contributed by all coupling reactions: i.e., an intrinsic tradeoff between global elasticity and thermodynamic cost. 
% : 
% \begin{align} \label{eq:SI_n_EPR_StrongLimit}
%         & \sigma_\cpl^{(n)} \simeq \sum_{i>j} k_\cpl^{-1} 
%         {\cdots\times}
%         \sqrt{\tilde{e}^+_i\tilde{e}^-_i}
%         \sqrt{\tilde{e}^+_j\tilde{e}^-_j}. 
% \end{align}

The above results are consistent with the case of $n=2$ in the main text.

\begin{table*}[tbh]
    \centering
    \caption{Order estimation in the case of ATP–GTP coupling. % (see also Supplementary Note~\ref{sec:order}).
    }
    \label{table:order}
    \begin{tabular}{c|c|c|c}
    Symbol & Description & Estimated value & Refs. \\ \hline 
    $\ATP_\tot$ & Pool size of ATP 
        & $1\text{--}5$ [mM] 
        & \cite{danchin1984metabolic,jewett2009continued,deng2021measuring,albe1990cellular} \\
    $\GTP_\tot$ & Pool size of GTP 
        & $0.1\text{--}1$ [mM] 
        & \cite{danchin1984metabolic,jewett2009continued,deng2021measuring} \\
    $\kc$ & Rate constant for the coupling reaction 
        & {$\sim10^4$ {${\rm [M^{-1}s^{-1}]}$} } 
        & \cite{zala2017advantage,tokarska2008nucleoside,pollack2002suspected} \\
    $\kappa^\pm_{\rm ATP}$ % , $\kappa_{\rm ADP}$ 
    & Effective rate constant for the driving reaction ${\rm ADP}\rightleftharpoons{\rm ATP}$
        & $\sim10^1$ {${\rm [s^{-1}]}$} 
        & \cite{deng2021measuring,albe1990cellular} \\
    $\Gamma_{\rm ATP}$ & $\ATP/[{\rm ADP}]$ ratio 
        & $1\text{--}10^2$ 
        & \cite{traut1994physiological,tantama2013imaging} \\
    $\Gamma_{\rm GTP}$ & $\GTP/[{\rm GDP}]$ ratio
        &  $1\text{--}10^2$
        & \cite{colombo1998involvement,traut1994physiological} \\
    $\rAB$ & $\ATP_\tot/\GTP_\tot$ ratio 
        &  $1\,\text{--}\;{>\,}10$
        &  \cite{danchin1984metabolic}
    \end{tabular}
\end{table*}

% {\color[cmyk]{0,.75,.90,0}
{\color{black}
\section{Relationship with metabolic control analysis}\label{sec:MCA}
The dimensionless elasticity $e^\pm_{XY}$, defined in Eq.~\eqref{eq:elasticity_def}, is conceptually close to sensitivity measures in Metabolic Control Analysis (MCA). One key difference is that our elasticity concerns the susceptibility of a ratio of steady-state concentrations, not concentrations themselves.

In MCA, the concentration control coefficient (CCC) is defined as the dependence of a steady-state concentration of metabolite $X$ on the activity of enzyme $E_i$,
\begin{equation*}
    C_{E_i}^{X} := \frac{\partial \log [X]}{\partial \log [E_i]}.
\end{equation*}
Thus, it is not directly equal to the dimensionless elasticity in Eq.~\eqref{eq:elasticity_def},
\begin{equation*}
    e_{XY}^{\pm} := \pm \frac{\partial \log\bigl([X^\ast]/[X]\bigr)}{\partial \log \kappa_Y^{\pm}}.
\end{equation*}
Rather, they are related via
\begin{equation}
    e_{XY}^{\pm}
    = \pm\left(C_{\kappa_Y^{\pm}}^{X^\ast} - C_{\kappa_Y^{\pm}}^{X}\right),
\end{equation}
where $\kappa_Y^{\pm}$ plays the role of an effective activity parameter for the coarse-grained driving reaction of currency metabolite $Y$. 
For clarity, note that the ``elasticity'' in MCA is typically defined as the dependence of a reaction flux $J_i$ on metabolite $X$, $\epsilon_{X}^{J_i} := \frac{\partial \log J_i}{\partial \log [X]}$, which differs from CCC.

MCA provides fundamental theorems for CCCs, such as the summation theorem,
\begin{equation*}
    \sum_{i} C_{E_i}^{X} = 0,
\end{equation*}
and the connectivity theorems,
\begin{equation*}
    \sum_{i} C_{E_i}^{X}\,\epsilon_{X}^{J_i} = -1, 
    \qquad
    \sum_{i} C_{E_i}^{X}\,\epsilon_{Y}^{J_i} = 0 \quad (X\neq Y).
\end{equation*}
The summation theorem is somewhat related to our identity
$e_{XA}^{\pm}+e_{XB}^{\pm}=1$, which implies
\begin{equation*}
    e_{XA}^{+}+e_{XB}^{+}-\left(e_{XA}^{-}+e_{XB}^{-}\right)=0
    \quad (X=A,B).
\end{equation*}
From the perspective of MCA, the summation theorem yields the identical equality,
\begin{equation*}
    \sum_{E=\kappa_A^{+},\kappa_A^{-},\kappa_B^{+},\kappa_B^{-}} C_{E}^{X^\ast} + \sum_{E=\kappa_A^{+},\kappa_A^{-},\kappa_B^{+},\kappa_B^{-}} C_{E}^{X} = 0
    \quad (X=A,B).
\end{equation*}
By contrast, it is not straightforward to apply the connectivity theorem to the elasticity $e_{XY}^{\pm}=\pm\left(C_{\kappa_Y^{\pm}}^{X^\ast}-C_{\kappa_Y^{\pm}}^{X}\right)$.}

\bibliography{bibliography}% Produces the bibliography via BibTeX.

\end{document}